\documentclass[useAMS,usenatbib]{mn2e}

\usepackage{graphicx}
\usepackage{amsmath,amssymb}
\newcommand{\apj}{ApJ}                                      
\newcommand{\apjs}{ApJS}
\newcommand{\apjl}{ApJL}
\newcommand{\aap}{A{\&}A}

\newcommand{\aaps}{A{\&}AS}
\newcommand{\aapr}{A{\&}A Rev}
\newcommand{\mnras}{MNRAS}
\newcommand{\aj}{AJ}

\newcommand{\pasp}{PASP}

\newcommand{\nat}{Nature}

\newcommand{\chir}{\ensuremath{\chi_\nu^{\,2}}}                   % Reduced chi-squared symbol
\newcommand{\schir}{\ensuremath{\sqrt{\chi_\nu^{\,2}}}}                   % Reduced chi-squared symbol

\title[Physical properties of WASP-45 and WASP-46]{Physical properties of the planetary systems WASP-45 and WASP-46 from simultaneous multi-band photometry \thanks{Based on data collected by the MiNDSTEp collaboration with the Danish 1.54 m telescope, and on data observed with the NTT (under program number 088.C-0204(A)), 2.2m and Euler-Swiss Telescope all located at the ESO La Silla Observatory.}}
\author[Ciceri et al.
]{S. Ciceri$^{1}$\thanks{E-mail: ciceri@mpia.de}, L. Mancini $^{1,2}$, J. Southworth$^{3}$, M. Lendl$^{4,5}$, J. Tregloan-Reed$^{6,3}$,
\newauthor R. Brahm$^{7,8}$, G. Chen$^{9,10}$, G. D'Ago$^{11,12,22}$, M. Dominik$^{13}$, R. Figuera Jaimes$^{13,14}$,
\newauthor P. Galianni$^{13}$, K. Harps{\o}e$^{15}$, T.~C. Hinse$^{16}$, U.~G. J{\o}rgensen$^{15}$, D. Juncher$^{15}$,
\newauthor H. Korhonen$^{15,17}$, C. Liebig$^{13}$, M. Rabus$^{1,7}$, A.~S. Bonomo$^{2}$, K. Bott$^{18,19}$,
\newauthor Th. Henning$^{1}$, A. Jord\'{a}n$^{7,8}$, A. Sozzetti$^{2}$, K.~A. Alsubai$^{20}$, J.~M. Andersen$^{15,21}$,
\newauthor D. Bajek$^{13}$, V. Bozza$^{12,22}$, D.~M. Bramich$^{20}$, P. Browne$^{13}$, S. Calchi Novati$^{23,11,12}$\footnotemark[1]\thanks{Sagan visiting fellow},
\newauthor Y. Damerdji$^{24}$, C. Diehl$^{25,26}$, A. Elyiv$^{24,27,28}$, E. Giannini$^{25}$, S-H. Gu$^{29,30}$,
\newauthor M. Hundertmark$^{15}$, N. Kains$^{31}$, M. Penny$^{32,15}$, A. Popovas$^{15}$, S. Rahvar$^{33}$, 
\newauthor G. Scarpetta$^{11,12,22}$, R.~W. Schmidt$^{25}$, J. Skottfelt$^{35,15}$, C. Snodgrass$^{35,36}$, J. Surdej$^{24}$, 
\newauthor C. Vilela$^{3}$, X-B. Wang$^{29,30}$ and O. Wertz$^{24}$\\
\newauthor$  $ \\
$^{1}$Max Planck Institute for Astronomy, K\"{o}nigstuhl 17, 69117 -- Heidelberg, Germany \\
$^{2}$INAF -- Osservatorio Astrofisico di Torino, via Osservatorio 20, 10025 -- Pino Torinese, Italy\\
$^{3}$Astrophysics Group, Keele University, Staffordshire, ST5 5BG, UK\\
$^{4}$Austrian Academy of Sciences, Space Research Institute, Schmiedlstrasse 6, 8042 Graz, Austria
$^{5}$Observatoire de Gen\`{e}ve, Universit\'{e} de Gen\`{e}ve, Chemin des maillettes 51, 1290 Sauverny, Switzerland\\
$^{6}$NASA Ames Research Center, Moffett Field, CA 94035, USA\\
$^{7}$Instituto de Astrof\'{i}sica, Pontificia Universidad Cat\'{o}lica de Chile, Av. Vicu\~{n}a Mackenna 4860, 7820436 -- Macul, Santiago, Chile\\ %
$^{8}$Millennium Institute of Astrophysics, Av. Vicu\~{n}a Mackenna 4860, 7820436 -- Macul, Santiago, Chile\\
$^{9}$Key Laboratory of Planetary Sciences, Purple Mountain Observatory, Chinese Academy of Sciences, Nanjing 210008, China\\
$^{10}$Instituto de Astrof\'{i}sica de Canarias, E-38205 La Laguna, Tenerife, Spain\\
$^{11}$International Institute for Advanced Scientific Studies (IIASS), Via G. Pellegrino 19, 84019 -- Vietri Sul Mare (SA), Italy\\    
$^{12}$Department of Physics ``E.\,R. Caianiello'', University of Salerno, Via Giovanni Paolo II 132, Fisciano (SA) -- 84084\\
$^{13}$SUPA, University of St Andrews, School of Physics \& Astronomy, North Haugh, St Andrews, Fife KY16 9SS, UK\\
$^{14}$European Southern Observatory, Karl-Schwarzschild Stra{\ss}e 2, 85748 -- Garching bei M\"{u}nchen, Germany\\
$^{15}$Niels Bohr Institute \& Centre for Star and Planet Formation, University of Copenhagen, {\O}ster Voldgade 5, 1350 -- Copenhagen K, Denmark\\
$^{16}$Korea Astronomy and Space Science Institute, 305-348 Daejeon, Republic of Korea\\
$^{17}$Finnish Centre for Astronomy with ESO (FINCA), University of Turku, V\"{a}is\"{a}l\"{a}ntie 20, FI-21500 Piikki\"{o}, Finland\\
$^{18}$School of Physics, UNSW Australia, NSW 2052, Australia\\
$^{19}$Australian Centre for Astrobiology, UNSW Australia, NSW 2052, Australia\\
$^{20}$Qatar Environment and Energy Research Institute, Qatar Foundation, Tornado Tower, Floor 19, PO Box 5825, Doha, Qatar\\
$^{21}$Department of Astronomy, Boston University, 725 Commonwealth Avenue, Boston, MA 02215, USA\\
$^{22}$Istituto Nazionale di Fisica Nucleare, Sezione di Napoli, I-80126 Napoli, Italy\\
$^{23}$NASA Exoplanet Science Institute, MS 100-22, California Institute of Technology, Pasadena CA 91125\\
$^{24}$Institut d'Astrophysique et de G\'{e}ophysique, Universit\'{e} de Li\`{e}ge, All\'{e}e du 6 Ao\^{u}t 17, B\^{a}t. B5C, Li\`{e}ge 1, 4000 Belgium\\
$^{25}$Astronomisches Rechen-Institut, Zentrum f\"{u}r Astronomie, Universit\"{a}t Heidelberg, Mönchhofstra{\ss}e 12-14, D-69120 Heidelberg, Germany\\
$^{26}$Hamburger Sternwarte, Universit\"{a}t Hamburg, Gojenbergsweg 112, D-21029 Hamburg, Germany\\
$^{27}$Dipartimento di Fisica e Astronomia, Università di Bologna, Viale Berti Pichat 6/2, I-40127 Bologna, Italy\\
$^{28}$Main Astronomical Observatory, Academy of Sciences of Ukraine, vul. Akademika Zabolotnoho 27, UA-03680 Kyiv, Ukraine\\
$^{29}$Yunnan Observatories, Chinese Academy of Sciences, Kunming 650011, China\\
$^{30}$Key Laboratory for the Structure and Evolution of Celestial Objects, Chinese Academy of Sciences, Kunming 650011,China\\
$^{31}$Space Telescope Science Institute, 3700 San Martin Drive, Baltimore, MD 21218, USA\\
$^{32}$Department of Astronomy, Ohio State University, McPherson Laboratory, 140 West 18th Avenue, Columbus, OH 43210-1173, USA\\
$^{33}$Department of Physics, Sharif University of Technology, PO Box 11155-9161 Tehran, Iran\\
$^{34}$Centre for Electronic Imaging, Dept. of Physical Sciences, The Open University, Milton Keynes MK7 6AA, UK\\
$^{35}$Planetary and Space Sciences, Department of Physical Sciences, The Open University, Milton Keynes, MK7 6AA, UK\\
$^{36}$Max Planck Institute for Solar System Research, Justus-von-Liebig-Weg 3, D-37077 G\"{o}ttingen, Germany}

\begin{document}

\date{2015 November 16}

\pagerange{\pageref{firstpage}--\pageref{lastpage}} \pubyear{2015}

\maketitle

\label{firstpage}

\begin{abstract}

Accurate measurements of the physical characteristics of a large number of exoplanets are useful to strongly constrain theoretical models of planet formation and evolution, which lead to the large variety of exoplanets and planetary-system configurations that have been observed. 
We present a study of the planetary systems WASP-45 and WASP-46, both composed of a main-sequence star and a close-in hot Jupiter, based on 29 new high-quality light curves of transits events.
In particular, one transit of WASP-45\,b and four of WASP-46\,b were simultaneously observed in four optical filters, while one transit of WASP-46\,b was observed with the NTT obtaining a precision of $0.30$ mmag with a cadence of roughly three minutes. We also obtained five new spectra of WASP-45 with the FEROS spectrograph.
We improved by a factor of four the measurement of the radius of the planet WASP-45\,b, and found that WASP-46\,b is slightly less massive and smaller than previously reported. Both planets now have a more accurate measurement of the density ($0.959 \pm 0.077\, \rho_{\mathrm{Jup}}$ instead of $0.64 \pm 0.30\, \rho_{\mathrm{Jup}}$ for WASP-45\,b, and $1.103\pm0.052\,\rho_{\mathrm{Jup}}$ instead of $0.94\pm0.11\,\rho_{\mathrm{Jup}}$ for WASP-46\,b).
We tentatively detected radius variations with wavelength for both planets, in particular in the case of WASP-45\,b we found a slightly larger absorption in the redder bands than in the bluer ones. 
No hints for the presence of an additional planetary companion in the two systems were found either from the photometric or radial velocity measurements.

\end{abstract}


%  Sect. 2
%%%%%%%%%%%%%%%%%%%%%%%%%%%%%%%%%%%%%%%%%%%%%%%%%%%%%%
\section{Introduction}
\label{sec_1}
%%%%%%%%%%%%%%%%%%%%%%%%%%%%%%%%%%%%%%%%%%%%%%%%%%%%%%

The possibility to obtain detailed information on extrasolar planets, using different techniques and methods, has revealed some unexpected properties that are still challenging astrophysicists. One of the very first was the discovery of Jupiter-like planets on very tight orbits, which are labelled hot Jupiters, and the corresponding inflation-mechanism problem (\citealp{baraffe2014}, and references therein). To find the answers to this and other open questions, it is important to have a proper statistical sample of exoplanets, whose physical and orbital parameters are accurately measured.

One class of extrasolar planets, those which transit their host stars, has lately seen a large increase in the number of its known members. This achievement has been possible thanks to systematic transit-survey large programs, performed both from ground (HATNet: \citealp{bakos:2004}; TrES: \citealp{alonso2004}; XO: \citealp{mcCullough:2005}; WASP: \citealp{pollacco:2006}; KELT: \citealp{pepper:2007}; MEarth: \citealp{charbonneau:2009}; QES: \citealp{alsubai2013}; HATSouth: \citealp{bakos:2013}), and from space (CoRoT: \citealp{barge:2008}, \emph{Kepler}: \citealp{borucki2011}).

The great interest in transiting planets lies in the fact that it is possible to measure all their main orbital and physical parameters with standard astronomical techniques and instruments. From photometry we can estimate the period, the relative size of the planet and the orbital inclination, whilst precise spectroscopic measurements provide a lower limit for the mass of the planet (but knowing the inclination from photometry the precise mass can be calculated) and the eccentricity of its orbit. Unveiling the bulk density of the planets allows the imposition of constraints on, or differentiation between, the diverse formation and migration theories which have been advanced (see \citealp{kley2012, baruteau2014} and references therein).

Furthermore, transiting planets allow astronomers to investigate their atmospheric composition, when observed in transit or occultation phases (e.g.\ \citealp{charbonneau2002, richardson2003}).
However, is also important to stress that, besides instrumental limitations, the characterisation of planets' atmosphere is made difficult by the complexity of their nature. Retrieving the atmospheric chemical composition may be hindered by the presence of clouds, resulting in a featureless spectrum (e.g.\ GJ\,1214\,b, \citealp{kreidberg:2014}).

In this work, we focus on two transiting exoplanet systems: WASP-45 and WASP-46. Based on new photometric and spectroscopic data, we review their physical parameters and probe the atmospheres of their planets.
% Sect. 1.1
\subsection{WASP-45}
\label{sec_1.1}
WASP-45 is a planetary system discovered within the SuperWASP survey by \cite{anderson2012} (A12 hereafter). The light curve of WASP-45, shows a periodic dimming (every $P=3.126$\,days) due to the presence of a hot Jupiter (radius $1.16^{+0.28}_{-0.14}\, R_{\mathrm{Jup}}$ and mass $1.007\pm0.053\, M_{\mathrm{Jup}}$) that transits the stellar disc. The host (mass of $0.909\pm0.060\, M_{\sun}$ and radius of $0.945^{+0.087}_{-0.071}\, R_{\sun}$) is a K2\,V star with a higher metallicity than the Sun ([Fe/H] $=$ $+0.36\pm 0.12$). The study of the Ca\,{\small II}\, H+K lines in the star's spectra revealed weak emission lines, which indicates a low chromospheric activity.

% Sect. 1.2
\subsection{WASP-46}
\label{sec_1.2}
WASP-46 is a G6\,V type star (mass $0.956\pm0.034\, M_{\sun}$, radius $0.917\pm0.028\, R_{\sun}$ and [Fe/H] $=$ $-0.37\pm0.13$). %\cite{anderson2012} 
A12 discovered a hot Jupiter (mass $2.101\pm0.073\, M_{\mathrm{Jup}}$ and radius $1.310\pm0.051\,R_{\mathrm{Jup}}$) that orbits the star every 1.430 days on a circular orbit. As in the case of WASP-45, weak emission visible in the Ca\,{\small II}\, H+K lines indicates a low stellar activity. Moreover, the WASP light curve shows a photometric modulation, which allowed the measurement of the rotation period of the star, and a gyrochronological age of the system of 1.4\,Gyr. By observing the secondary eclipse of the planet, \cite{chen2014} detected the emission from the day-side atmosphere finding brightness temperatures consistent with a low heat redistribution efficiency.

%  Sect. 2
%%%%%%%%%%%%%%%%%%%%%%%%%%%%%%%%%%%%%%%%%%%%%%%%%%%%%%
\section[]{Observations and data reduction}
\label{sec_2}
%%%%%%%%%%%%%%%%%%%%%%%%%%%%%%%%%%%%%%%%%%%%%%%%%%%%%%
A total of 29 light curves of 14 transit events of WASP-45\,b and WASP-46\,b, and five spectra of WASP-45 were obtained using four telescopes located at the ESO La Silla observatory, Chile. The details of the photometric observations are reported in Table\,\ref{ObsLog}.

The photometric observations were all performed with the telescopes in auto-guided mode and out of focus to minimise the sources of noise and increase the signal to noise ratio (S/N). By spreading the light of each single star on many more pixels of the CCD, it is then possible to utilise longer exposure times without incurring saturation.
Atmospheric variations, change in seeing, or tracking imprecision lead to changes in the position and/or size of the point spread function (PSF) on the CCD and, according to the different response of each single pixel, spurious noise can be introduced in the signal. By obtaining a doughnut-shaped PSF that covers a circular region with a diameter from roughly 15 to 30 pixels, the small variations in position are averaged out and have a much lower effect on the photometric precision \citep{southworth2009}.

% Table 1
\begin{table*}
\centering %
\caption{Details of the transit observations presented in this work. $N_{\rm obs}$ is the number of observations, $T_{\rm exp}$ is the exposure time, $T_{\rm obs}$ is the observational cadence, and `Moon illum.' is the fractional illumination of the Moon at the midpoint of the transit. The triplets of numbers in the \textit{Aperture radii} column correspond to the inner radius of the circular aperture (star) and outer radii of the ring (sky) selected for the photometric measurements.  $^{a}$ The aperture radii of the Euler-Swiss light curves are referred to the target aperture and are expressed in arcsec.}
\scriptsize  %
\begin{tabular}{lccccccccccc}
\hline %
\hline %
Telescope & Date of   & Start time & End time  &$N_{\rm obs}$ & $T_{\rm exp}$ & $T_{\rm obs}$ & Filter & Airmass & Moon & Aperture   & Scatter \\
          & first obs &    (UT)    &   (UT)    &              & (s)           & (s)           &        &         &illum.& radii (px) & (mmag)  \\
\hline %
\multicolumn{10}{l}{WASP-45:} \\
Eul\,1.2m  & 2011 Nov 07 & 02:12 & 05:57 & 197 &    50     &  69  & Gunn $r$   & 1.00 $\to$ 1.51 & 89\% & 6.0  $^{a}$& 0.91 \\[2pt]
Dan\,1.54m & 2012 Sep 02 & 05:05 & 08:46 & 122 &   100     & 106  & Bessel $R$ & 1.15 $\to$ 1.51 & 93\% & 17,60,75  & 0.55 \\[2pt]
MPG\,2.2m  & 2012 Oct 15 & 23:40 & 03:58 & 238 &    40     &  65  & Sloan $g'$ & 1.46 $\to$ 1.01 &  1\% & 23,70,85  & 0.85 \\
MPG\,2.2m  & 2012 Oct 15 & 23:40 & 03:58 & 240 &    40     &  65  & Sloan $r'$ & 1.46 $\to$ 1.01 &  1\% & 23,55,70  & 0.65 \\
MPG\,2.2m  & 2012 Oct 15 & 23:40 & 03:58 & 238 &    40     &  65  & Sloan $i'$ & 1.46 $\to$ 1.01 &  1\% & 21,45,65  & 0.83 \\
MPG\,2.2m  & 2012 Oct 15 & 23:40 & 03:58 & 241 &    40     &  65  & Sloan $z'$ & 1.46 $\to$ 1.01 &  1\% & 17,50,70  & 0.75 \\[2pt]
Dan\,1.54m & 2013 Jul 24 & 07:47 & 10:43 & 100 &   100     & 106  & Bessel $R$ & 1.15 $\to$ 1.34 & 97\% & 13,55,70  & 0.62 \\
\hline %
\multicolumn{10}{l}{WASP-46:} \\
Eul\,1.2m  & 2011 Jun 10 & 03:40 & 07:11 & 117 &    90     & 108  & Gunn $r$   & 1.95 $\to$ 1.17 & 63\% & 5.2 $^{1}$& 1.26 \\
Eul\,1.2m  & 2011 Sep 01 & 02:47 & 06:37 &  72 &   170     & 180  & Gunn $r$   & 1.12 $\to$ 1.36 & 14\% & 4.3 $^{1}$& 0.88 \\ [2pt]
NTT\,3.58m & 2011 Oct 24 & 00:42 & 05:24 &  85 &   150     & 176  & Gunn $g$   & 1.14 $\to$ 2.21 & 10\% & 52,80,100 & 0.30 \\ [2pt]
MPG\,2.2m  & 2012 Jul 03 & 04:19 & 10:29 & 116 &   115     & 145  & Sloan $g'$ & 1.12 $\to$ 1.39 &100\% & 20,65,80  & 0.56 \\
MPG\,2.2m  & 2012 Jul 03 & 04:19 & 10:29 & 121 &   115     & 145  & Sloan $r'$ & 1.12 $\to$ 1.39 &100\% & 22,65,80  & 0.64 \\
MPG\,2.2m  & 2012 Jul 03 & 04:19 & 10:29 & 120 &   115     & 145  & Sloan $i'$ & 1.12 $\to$ 1.39 &100\% & 22,65,80  & 0.56 \\
MPG\,2.2m  & 2012 Jul 03 & 04:19 & 10:29 & 119 &   115     & 145  & Sloan $z'$ & 1.12 $\to$ 1.39 &100\% & 23,65,80  & 0.62 \\[2pt]
Dan\,1.54m & 2012 Sep 24 & 04:24 & 08:03 &  97 &   120     & 131  & Bessel $R$ & 1.14 $\to$ 1.95 & 66\% & 11,65,80  & 1.52 \\[2pt]
MPG\,2.2m  & 2012 Oct 17 & 01:03 & 05:55 & 106 &    90     & 116  & Sloan $g'$ & 1.13 $\to$ 2.27 &  4\% & 25,90,105 & 0.80 \\
MPG\,2.2m  & 2012 Oct 17 & 01:03 & 05:55 & 107 &    90     & 116  & Sloan $r'$ & 1.13 $\to$ 2.27 &  4\% & 25,90,105 & 0.75 \\
MPG\,2.2m  & 2012 Oct 17 & 01:03 & 05:55 & 109 &    90     & 116  & Sloan $i'$ & 1.13 $\to$ 2.27 &  4\% & 23,90,100 & 0.81 \\
MPG\,2.2m  & 2012 Oct 17 & 01:03 & 05:55 & 108 &    90     & 116  & Sloan $z'$ & 1.13 $\to$ 2.27 &  4\% & 23,90,105 & 0.85 \\[2pt]
MPG\,2.2m  & 2013 Apr 25 & 06:07 & 10:30 &  59 & 120 to 170& 215  & Sloan $g'$ & 2.18 $\to$ 1.15 &100\% & 24,65,80  & 0.70 \\
MPG\,2.2m  & 2013 Apr 25 & 06:07 & 10:30 &  59 & 120 to 170& 215  & Sloan $r'$ & 2.18 $\to$ 1.15 &100\% & 24,65,80  & 0.63 \\
MPG\,2.2m  & 2013 Apr 25 & 06:07 & 10:30 &  57 & 120 to 170& 215  & Sloan $i'$ & 2.18 $\to$ 1.15 &100\% & 25,65,80  & 0.90 \\
MPG\,2.2m  & 2013 Apr 25 & 06:07 & 10:30 &  57 & 120 to 170& 215  & Sloan $z'$ & 2.18 $\to$ 1.15 &100\% & 24,65,80  & 0.81 \\[2pt]
MPG\,2.2m  & 2013 Jun 17 & 05:36 & 09:01 &  83 & 130 to 70 & 137  & Sloan $g'$ & 1.28 $\to$ 1.12 & 55\% & 26,70,85  & 0.61 \\
MPG\,2.2m  & 2013 Jun 17 & 05:36 & 09:01 &  82 & 130 to 70 & 137  & Sloan $r'$ & 1.28 $\to$ 1.12 & 55\% & 26,70,85  & 0.64 \\
MPG\,2.2m  & 2013 Jun 17 & 05:36 & 09:01 &  84 & 130 to 70 & 137  & Sloan $i'$ & 1.28 $\to$ 1.12 & 55\% & 28,65,80  & 0.70 \\
MPG\,2.2m  & 2013 Jun 17 & 05:36 & 09:01 &  81 & 130 to 70 & 137  & Sloan $z'$ & 1.28 $\to$ 1.12 & 55\% & 28,70,85  & 0.85 \\[2pt]
Dan\,1.54m & 2013 Aug 06 & 07:19 & 10:28 &  99 &   100     & 115  & Bessel $R$ & 1.12 $\to$ 1.53 &  1\% & 11,65,80  & 0.68 \\
Dan\,1.54m & 2013 Aug 28 & 04:06 & 08:39 & 132 &   100     & 125  & Bessel $R$ & 1.15 $\to$ 1.43 & 42\% & 12,65,80  & 0.88 \\
\hline
\end{tabular}
\label{ObsLog}%
\end{table*}

% Figure 01
\begin{figure}
\centering
\includegraphics[width=\columnwidth]{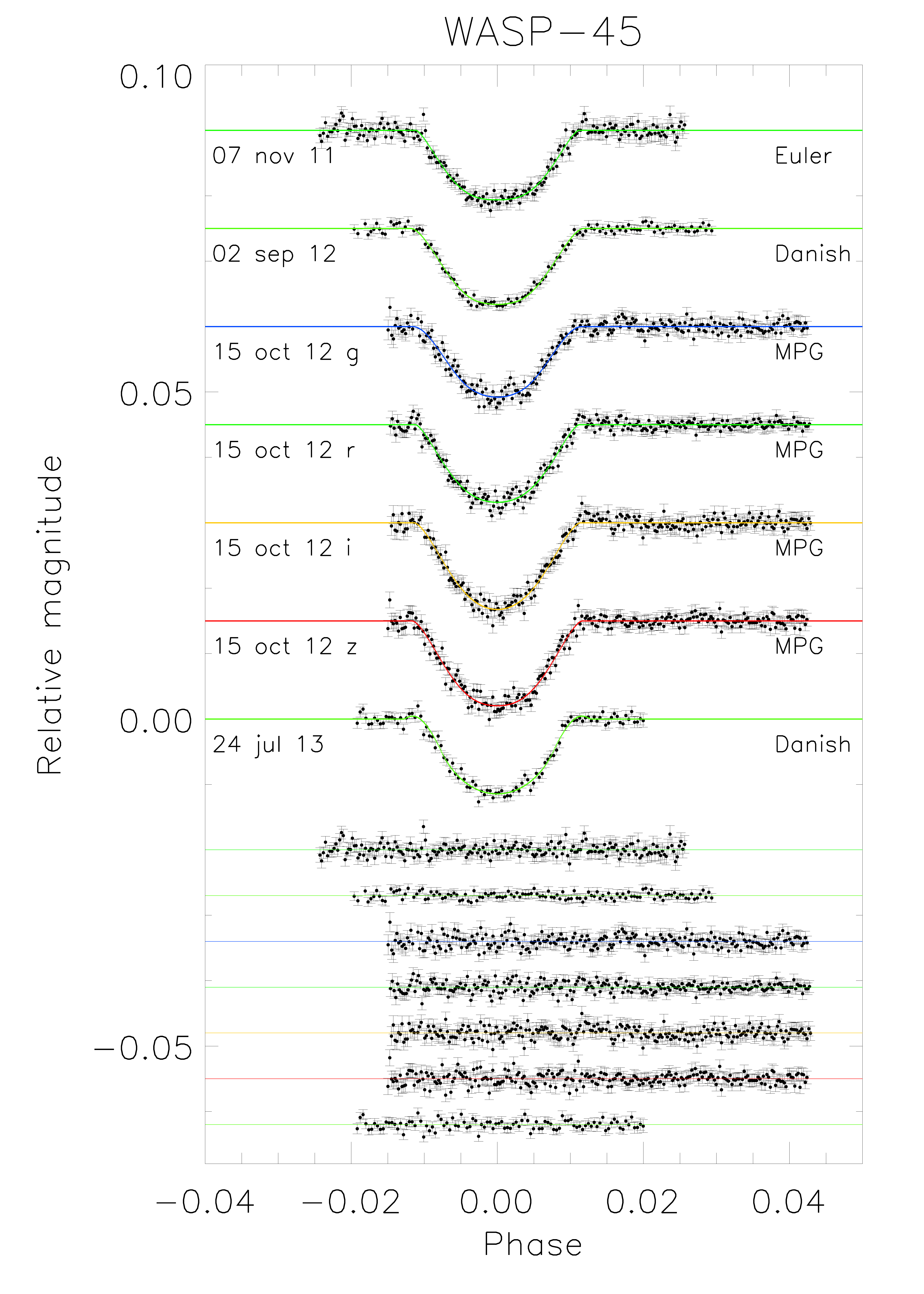}
\caption{Phased light curves corresponding to four transits of WASP-45\,b. The date and the telescope used for each transit event are indicated close to the corresponding light curve. Each light curve is shown with its best fit obtained with {\sc jktebop}. The residuals relative to each fit are displayed at the base of the figure in the same order as the light curves. The curves are shifted along the y axis for clarity.}
\label{w45_lc}
\end{figure}

% Figure 02
\begin{figure}
\centering
\includegraphics[width=\columnwidth]{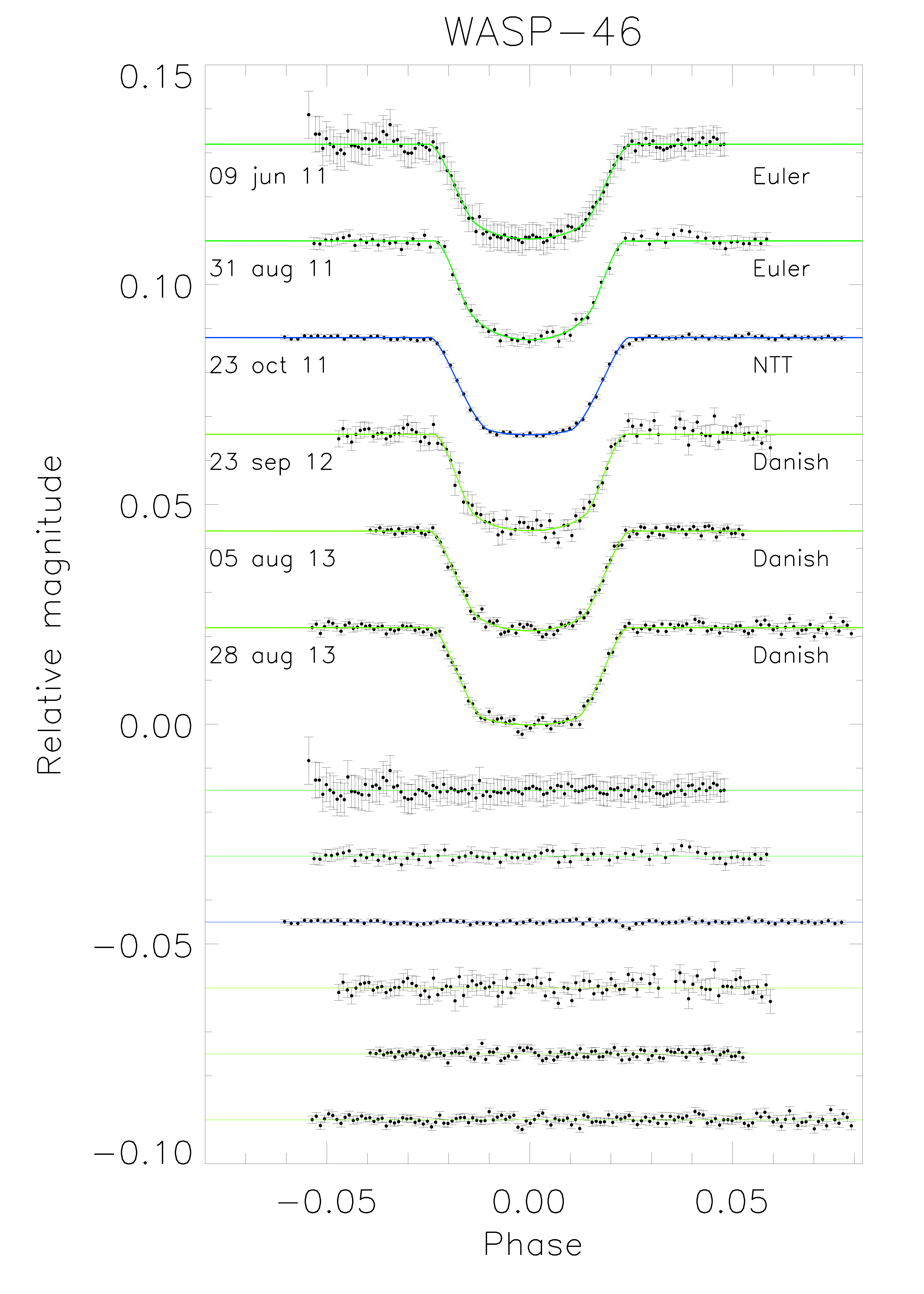}
\caption{Phased light curves of the transits of WASP-46\,b, observed with the Euler, NTT, and Danish telescopes, compared with the best-fitting curves given by {{\sc jktebop}}. The date and the telescope used for each transit event are indicated close to the corresponding light curve. The residuals relative to each fit are displayed at the base of the figure in the same order as the light curves.The curves are shifted along the y axis for clarity.}
\label{w46_lc}
\end{figure}

% Figure 03
\begin{figure}
\centering
\includegraphics[width=\columnwidth]{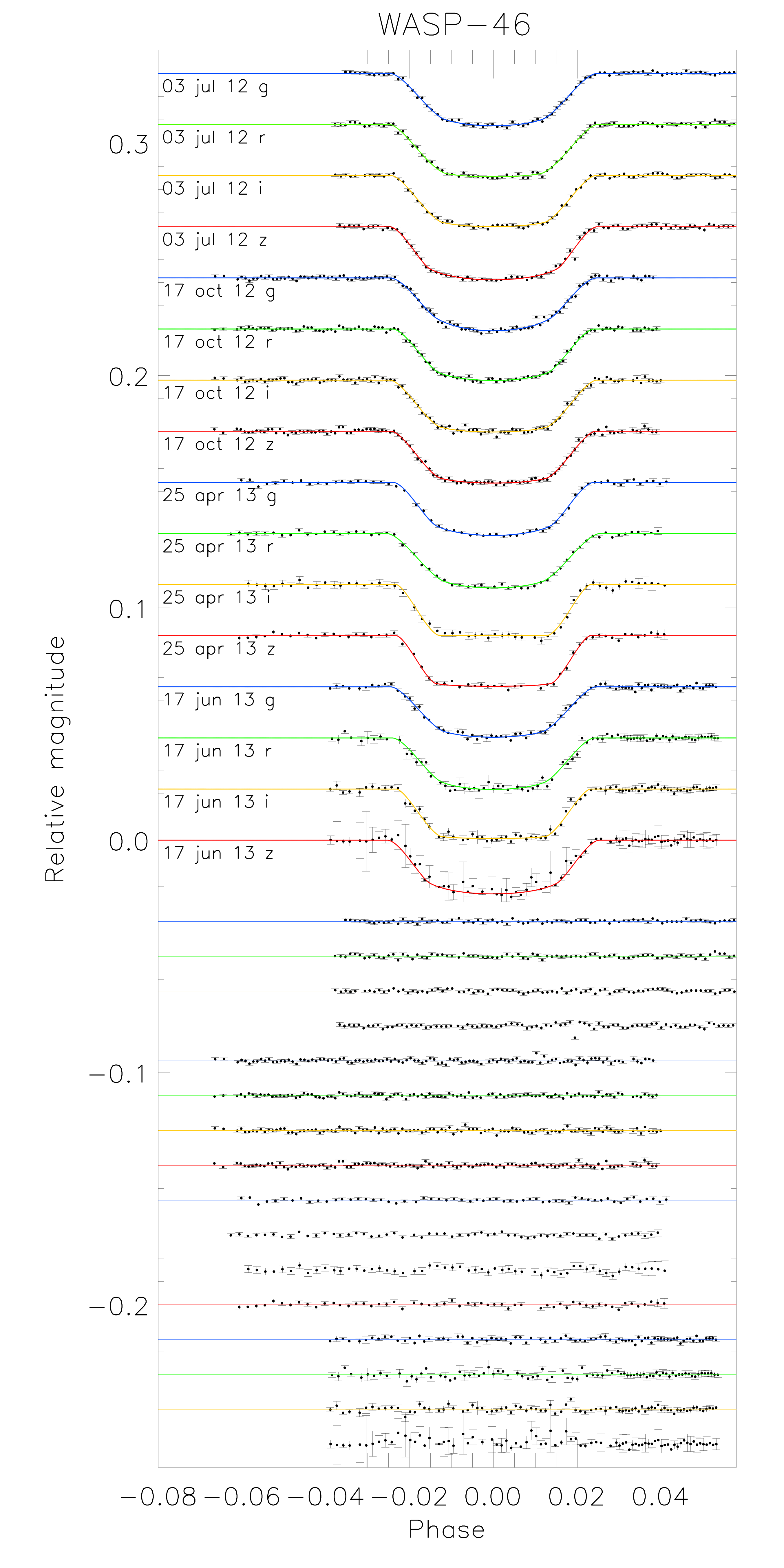}
\caption{Phased light curves of four transits of WASP-46 observed with GROND. Each transit was observed simultaneously in four optical bands. As in Figs.\ \ref{w45_lc} and \ref{w46_lc} the light curves are compared with the best-fitting curves given by {{\sc jktebop}}, and the residuals from the fits are displayed at the bottom in the same order as the light curves.The curves are shifted along the y axis for clarity.}
\label{w46_grond}
\end{figure}

%%%%%%%%%%%%%%%%%%%%%%%%%%%%%%%%%%%%%%%%%%%%%%%%%%%%%%
\subsubsection*{1.2\,m Euler-Swiss Telescope}
%%%%%%%%%%%%%%%%%%%%%%%%%%%%%%%%%%%%%%%%%%%%%%%%%%%%%%
The imager of the 1.2m Euler-Swiss telescope, EulerCam, is a  $4000 \times 4000$ e2v CCD with a field of view (FOV) of $15.7^{\prime} \times 15.7^{\prime}$, yielding a resolution of 0.23 arcsec per pixel. Since its installation in 2010, EulerCam has been used intensively for photometric follow-up observations of planet candidates from the WASP survey (e.g. \citealp{hellier2011, lendl2014}), as well as for the atmospheric study of highly irradiated giant planets \citep{lendl2013}. We observed one transit of WASP-45 b and two transits of WASP-46 b with EulerCam between June and September 2011 using a Gunn \textit{r'} fiter.

%%%%%%%%%%%%%%%%%%%%%%%%%%%%%%%%%%%%%%%%%%%%%%%%%%%%%%
\subsubsection*{3.58\,m NTT}
%%%%%%%%%%%%%%%%%%%%%%%%%%%%%%%%%%%%%%%%%%%%%%%%%%%%%%
One transit of WASP-46\,b was observed with the New Technology Telescope (NTT) on the 23 October 2011. The telescope has a primary mirror of 3.58\,m and is equipped with an active optics system. The EFOSC2 instrument (ESO Faint Object Spectrograph and Camera 2), mounted on its Nasmyth\,B focus, was utilised. The field of view of its Loral/Lesser camera is $4.1^{\prime} \times 4.1^{\prime}$ with a resolution of 0.12 arcsec per pixel. During the observation of the whole transit, only a small region of the CCD, including the target and some reference stars, was read out in order to diminish the readout time and increase the sampling. The filter used was a Gunn $g$ (ESO \#782).

%%%%%%%%%%%%%%%%%%%%%%%%%%%%%%%%%%%%%%%%%%%%%%%%%%%%%%
\subsubsection*{1.54\,m Danish Telescope}
%%%%%%%%%%%%%%%%%%%%%%%%%%%%%%%%%%%%%%%%%%%%%%%%%%%%%%
Two transits of WASP-45\,b and three of WASP-46\,b were observed with the 1.54\,m Danish Telescope, using the DFOSC (Danish Faint Object Spectrograph and Camera) instrument mounted at the Cassegrain focus. The instrument, now used exclusively for imaging, has a CCD with a FOV $13.7^{\prime} \times 13.7^{\prime}$ and a resolution of 0.39 arcsec per pixel. The CCD was windowed and a Bessell $R$ filter was used for all the transits.

%%%%%%%%%%%%%%%%%%%%%%%%%%%%%%%%%%%%%%%%%%%%%%%%%%%%%%
\subsubsection*{2.2\,m MPG Telescope - GROND}
%%%%%%%%%%%%%%%%%%%%%%%%%%%%%%%%%%%%%%%%%%%%%%%%%%%%%%
The 2.2\,m MPG Telescope holds in its Coud\'{e}-like focus the GROND (Gamma Ray Optical Near-infrared Detector) instrument (\citealp{greiner2008}). GROND is a seven channel imager capable of performing simultaneous observations in four optical bands ($g^{\prime}$, $r^{\prime}$, $i^{\prime}$, $z^{\prime}$, similar to Sloan filters) and three near-infrared (NIR) bands ($J$, $H$, $K$). The light, split into different paths using dichroics, reaches two different sets of cameras. The optical cameras have $2048 \times 2048$ pixels with a resolution of 0.16 arcsec per pixel. The NIR cameras have a lower resolution 0.60 arcsec per pixel, but a larger FOV of $10^{\prime} \times 10^{\prime}$ (almost double the optical ones).  The primary goal of GROND is the detection and follow-up of the optical/NIR counterpart of gamma ray bursts, but it has already proven to be a great instrument to perform multicolour, simultaneous, photometric observations of planetary-transit events (e.g.\ \citealp{mancini2013b, southworth2015}). With this instrument, we observed one transit event of the planet WASP-45\,b and four of WASP-46\,b.

The exposure time must be the same for each optical camera, and is also partially constrained by the NIR exposure time chosen. We therefore decided to fix the exposure time to that optimising the $r'$-band counts (generally higher than in the other bands) in order to avoid saturation.

%%%%%%%%%%%%%%%%%%%%%%%%%%%%%%%%%%%%%%%%%%%%%%%%%%%%%
\subsubsection*{2.2\,m MPG Telescope - FEROS}
\label{subsec_2.5}
%%%%%%%%%%%%%%%%%%%%%%%%%%%%%%%%%%%%%%%%%%%%%%%%%%%%%%
The 2.2\,m MPG telescope also hosts FEROS (Fibre-fed Extended Range Optical Spectrograph). This \'echelle spectrograph covers a wide wavelength range of 370\,nm to 860\,nm and has an average resolution of $R=48\,000 \pm 4\,000$. The precision of the radial velocity (RV) measurements obtained with FEROS is good enough ($\geq 10$\,m\,s$^{-1}$) for detecting and confirming Jupiter-size exoplanets (e.g.\ \citealp{penev2013, jones2015}). Simultaneously to the science observations, we always obtained a spectrum of a ThAr lamp in order to have a proper wavelength calibration. Five spectra of WASP-45 were obtained with FEROS.

%%%%%%%%%%%%%%%%%%%%%%%%%%%%%%%%%%%%%%%%%%%%%%%%%%%%%%
\subsection{Data reduction}
\label{sec_2.1}
%%%%%%%%%%%%%%%%%%%%%%%%%%%%%%%%%%%%%%%%%%%%%%%%%%%%%%

For all photometric data, a suitable number of calibration frames, bias and (sky) flat-field images, were taken on the same day as the observations. Master bias and flat-field images were created by median-combining all the individual bias and flat-field images, and used to calibrate the scientific images.

With the exception of the EulerCam data, we then extracted the photometry from the calibrated images using a version of the aperture-photometry algorithm {\sc daophot} \citep{Stetson1987} implemented in the {\sc defot} pipeline \citep{southworth2014}. We measured the flux of the targets and of several reference stars in the FOVs, selecting those of similar brightness to the target and not showing any significant brightness variation due to intrinsic variability or instrumental effects. For each dataset, we tried different aperture sizes for both the inner and outer rings, and the final ones that we selected (see Table\,\ref{ObsLog}) were those that gave the lowest scatter in the out-of-transit (OOT) region. Light curves were then obtained by performing differential photometry using the reference stars in order to correct for non-intrinsic variations of the flux of the target, which are caused by atmospheric or air-mass changes. Also in this case we tried different combinations of multiple comparison stars and chose those that gave the lowest scatter in the OOT region. We noticed that the different options gave consistent transit shapes but had a small effect on the scatter of the points in the light curves. Finally, each light curve was obtained by optimising the weights of the chosen comparison stars.

The EulerCam photomety was also extracted using relative aperture photometry with the extraction being performed for a number of target and sky apertures, of which the best was selected based on the final lightcurve RMS. The selection of the reference stars was done iteratively, optimizing the scatter of the full transit lightcurves based on preliminary fits of transit shapes to the data at each optimization step. For further details, please see \cite{lendl2012}.

To properly compare the different light curves and avoid underestimation of the uncertainties assigned to each photometric point, we inflated the errors by multiplying them by the $\schir$ (defined as $\chir= \Sigma((x_i - x_{best})^{2}/ \sigma^{2})$/ DOF, where DOF is the number of degrees of freedom) obtained from the first fit of each light curve. We then took into account the possible presence of correlated noise or systematic effects using the $\beta$ approach (e.g.\ \citealp{gillon2006,winn2007}), with which we further enlarged the uncertainties. 

The NIR light curves observed with GROND were reduced following \cite{chen2014b}, by carefully subtracting the dark from each image and flat-fielding them, and correcting for the read-out pattern. No sky subtraction was performed since no such calibration files were available. Unfortunately, the quality of the data was not good enough to proceed with a detailed analysis of the transits.

%%%%%%%%%%%%%%%%%%%%%%%%%%%%%%%%%%%%%%%%%%%%%%%%%%%%%
\subsubsection*{FEROS spectra reduction}
\label{subsec_2.6}
%%%%%%%%%%%%%%%%%%%%%%%%%%%%%%%%%%%%%%%%%%%%%%%%%%%%%%
The spectra obtained with FEROS were extracted using a new pipeline written for \'echelle spectrographs, adapted for this instrument and optimised for the subsequent RV measurements (\citealp{jordan2014, brahm2015}). In brief, first a master-bias and a master-flat were constructed as the median of the frames obtained during the afternoon routine calibrations. The master-bias was subtracted from the science frames in order to account for the CCD intrinsic inhomogeneities, while the master-flat was used to find and trace all 39 \'echelle orders. The spectra of the target and the calibration ThAr lamp were extracted following \cite{Marsh1989}. The science spectrum was then calibrated in wavelength using the ThAr spectrum, and a barycentric correction was applied. In order to measure the RV of the star, the spectrum was cross-correlated with a binary mask chosen according to the spectral class of the target. For each \'echelle order a cross correlation function (CCF) was found and the RV measured by fitting a combined one, which is obtained as a weighted sum of all the CCF, with a Gaussian. The uncertainties on the RVs were calculated using empirical scaling relations from the width of the CCF and the mean S/N measured around 570\,nm. The RV measurements are reported in Table\,\ref{w45_rv}.

% Table 2
\begin{table}
\centering %
\caption{FEROS RV measurements of WASP-45.}
% \tiny %
\begin{tabular}{ccc}
\hline %
\hline %
Date of observation  &  RV    & err$_{\mathrm{RV}}$ \\
BJD-2400000          &  km\,s$^{-1}$ & km\,s$^{-1}$        \\
\hline %
56939.69704138 & 4.368 & 0.010 \\
56941.64378526 & 4.557 & 0.011 \\
56942.69196252 & 4.382 & 0.010 \\
57037.53926492 & 4.650 & 0.010 \\
57049.54110894 & 4.502 & 0.010 \\
\hline %
\end{tabular}
\label{w45_rv}%
\end{table}

%%%%%%%%%%%%%%%%%%%%%%%%%%%%%%%%%%%%%%%%%%%%%%%%%%%%%%
\section{Light curve analysis}
\label{sec_3}
%%%%%%%%%%%%%%%%%%%%%%%%%%%%%%%%%%%%%%%%%%%%%%%%%%%%%%

The light curve shape of a transit (its depth and duration) directly depends on values that describe the planet and its host star (e.g.\ \citealp{seager2003}).
In particular, by fitting the transit shape it is possible to obtain the measurement of the stellar and planetary relative radii, $ r_{*}=\frac{R_{*}}{a}$ and $ r_{b}=\frac{R_{b}}{a}$ (where $a$ is the semi-major axis of the orbit), the inclination of the planetary orbit with respect to the line of sight of the observer, $i$, and the time of the transit centre, $T_{0}$.

Using the {\sc jktebop}\footnote{The \textsc{jktebop} source code can be downloaded at {\tt http:// www.astro.keele.ac.uk/jkt/codes/jktebop.html}} code \citep[version 34,][and references therein]{southworth2013}, we separately fitted each light curve initially setting the fitted  parameters to the values published in the discovery paper.
The values for each parameter were then obtained through a Levenberg-Marquardt minimisation, while uncertainties were estimated by running Monte Carlo and residual-permutation algorithms \citep{southworth2008}. The coefficients of a second order polynomial were also fitted to account for instrumental and astrophysical trends possibly present in the light curves. In particular, $10\,000$ simulations for the Monte Carlo and Ndata-1 simulations for residual-permutation algorithm were run, and the larger of the two 1-$\sigma$ values were adopted as the final uncertainties. The {\sc jktebop} code is capable of simultaneously fitting light curves and RVs, and therefore giving also an estimation of the semi-amplitude $K$ and systemic velocity $\gamma_{\mathrm{sys}}$.

To properly constrain the planetary system's quantities we took into account the effect of the star's limb darkening (LD) while the planet is transiting the stellar disc. We applied a quadratic law to describe this effect, and used the LD coefficients provided by the stellar models of \cite{claret2000,claret2004} once the stellar atmospheric parameters were supplied (Table\,\ref{ld-coeff}). Each light curve was firstly fitted for the linear coefficient, while the quadratic one was perturbed during the Monte Carlo and residual-permutation algorithms in order to account for its uncertainty. Then we repeated the fitting process whilst keeping both the LD coefficients fixed.

% Table 3 ld coeff
\begin{table}
\centering
\caption{Stellar atmospheric parameters used to calculate the LD coefficients used to model the light curves.}
\begin{tabular}{lcc}
\hline
\hline
Parameter & WASP-45 & WASP-46 \\
\hline
$T_{\mathrm{eff}} (K)$             & 5100 & 5600  \\
$\log{g}$ (cm s$^{-2}$)            & 4.5 & 4.5  \\
$[\frac{\mathrm{Fe}}{\mathrm{H}}]$ & 0.5  & $-0.3$  \\
$V_{\mathrm{micro}}$ (km s$^{-1}$) & 2.0  & 2.0   \\
\hline
\end{tabular}
\label{ld-coeff}
\end{table}

Considering the discussion in A12, we fixed the eccentricity to zero for both the planetary systems. As an extra check, we used the Systemic Console 2 \citep{meschiari2009} to fit the RVs published in the discovery paper and those we observed with FEROS, obtaining a value consistent with $e=0$ (from the fit we obtained $e=0.041 \pm 0.043$ for WASP-45). All the light curves observed along with the best fit are shown in Fig.\ref{w45_lc} for WASP-45 and Figs. \ref{w46_lc} and \ref{w46_grond} for WASP-46.

%%%%%%%%%%%%%%%%%%%%%%%%%%%%%%%%%%%%%%%%%%%%%%%%%%%%%%
\subsection{New orbital ephemeris}%
\label{sec_3.1}
%%%%%%%%%%%%%%%%%%%%%%%%%%%%%%%%%%%%%%%%%%%%%%%%%%%%%%

\begin{figure*}
\centering
\includegraphics[width=\textwidth]{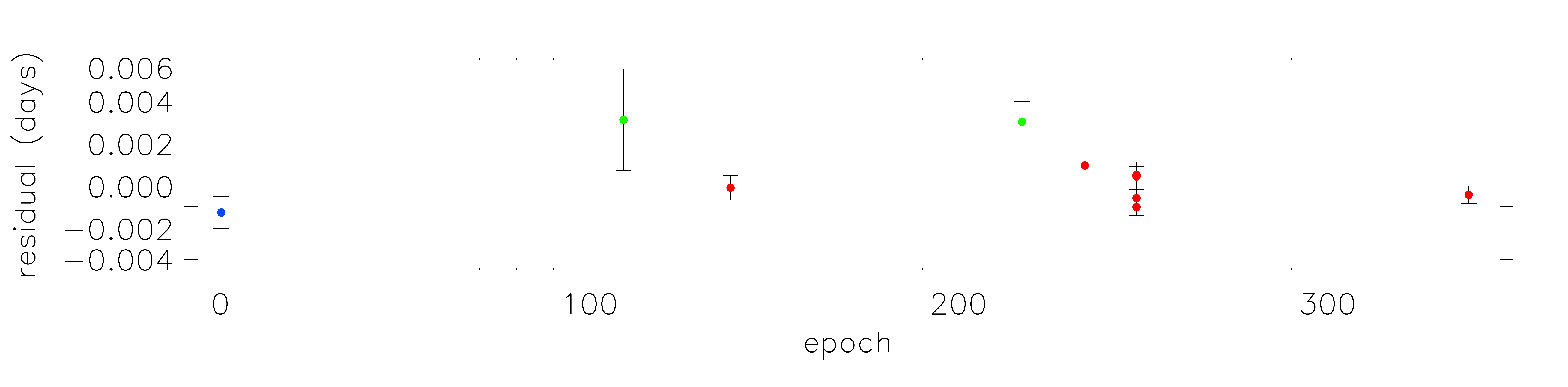}
\caption{Plot of the residuals of the timing of mid-transit of WASP-45 versus a linear ephemeris. The different colours of the points refer to the value from the discovery paper (blue), values obtained from the ETD catalogue (green), and our data (red).}
\label{oc_w45}
\end{figure*}

\begin{figure*}
\centering
\includegraphics[width=\textwidth]{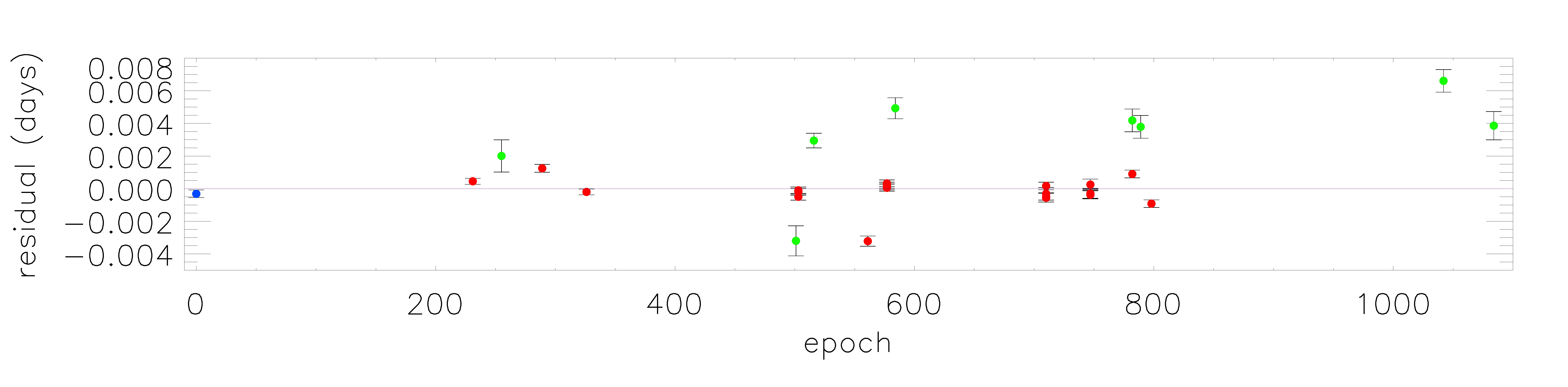}
\caption{Plot of the residuals of the timing of mid-transit of WASP-46 versus a linear ephemeris. The different colours of the points refer to the value from the discovery paper (blue), values obtained from the ETD catalogue (green), and our data (red).}
\label{oc_w46}
\end{figure*}

From the fit of each light curve, we obtained, among the other properties, accurate values of the mid-transit times. By also taking into account the values found from the discovery paper A12 and those from the Exoplanet Transit Database (ETD)\footnote{The database can be found on {\tt http://var2.astro.cz/ETD}} website, we refined the ephemeris values. In particular for WASP-46 we used only those light curves from the ETD catalogue that had a data quality index better than 3 and whose light curve didn't show evident deviation from a transit shape that could affect the $T_{0}$ measurement.
% Table 4
\begin{table}
\caption{Times of mid-transit point of WASP-45\,b and their residuals. References: (1)A12; (2) ETD; (3) Euler, this work; (4) GROND, this work; (5) Danish, this work.}
\centering
\footnotesize 
%\scriptsize  %
\begin{tabular}{lrrl}
\hline
\hline
~Time of minimum    & Epoch & Residual & Reference  \\
~BJD(TDB)$-2400000$ &       &   (JD)~~   &            \\
\hline
$ 55441.27000 \pm 0.00058 $ &   0 ~&-0.00128 & ~~~~(1) \\
$ 55782.01007 \pm 0.00235 $ & 109 ~& 0.00310 & ~~~~(2) \\
$ 55872.67006 \pm 0.00030 $ & 138 ~&-0.00011 & ~~~~(3) \\
$ 56119.62852 \pm 0.00081 $ & 217 ~& 0.00301 & ~~~~(2) \\
$ 56172.77422 \pm 0.00018 $ & 234 ~& 0.00094 & ~~~~(5) \\
$ 56216.54008 \pm 0.00029 $ & 248 ~& 0.00042 & ~~~~(4) $g$ \\
$ 56216.54153 \pm 0.00021 $ & 248 ~&-0.00103 & ~~~~(4) $r$ \\
$ 56216.54110 \pm 0.00024 $ & 248 ~&-0.00060 & ~~~~(4) $i$ \\
$ 56216.54001 \pm 0.00025 $ & 248 ~& 0.00049 & ~~~~(4) $z$ \\
$ 56497.88958 \pm 0.00026 $ & 338 ~&-0.00044 & ~~~~(5) \\
\hline
\end{tabular}
\label{T0_w45}
\end{table}
%
% Table 5
\begin{table}
\caption{Times of mid-transit point of WASP-46\,b and their residuals. References: (1) A12; (2) ETD; (3) Euler, this work; (4) NTT, this work; (5) GROND, this work; (6) Danish, this work.}
\centering
\footnotesize 
%\scriptsize  %
\begin{tabular}{lrrl}
\hline
\hline
~Time of minimum    & Epoch & Residual & Reference  \\
~BJD(TDB)$-2400000$ &       &   (JD)~~~   &            \\
\hline
$ 55392.31628 \pm 0.00020 $ &   0 ~&-0.00032& ~~~~(1) \\
$ 55722.73197 \pm 0.00013 $ & 231 ~& 0.00045& ~~~~(3) \\
$ 55757.06235 \pm 0.00098 $ & 255 ~& 0.00201& ~~~~(2) \\
$ 55805.69409 \pm 0.00020 $ & 289 ~& 0.00125& ~~~~(3) \\
$ 55858.61624 \pm 0.00011 $ & 326 ~&-0.00020& ~~~~(4) \\
$ 56108.92758 \pm 0.00091 $ & 501 ~&-0.00320& ~~~~(2) \\
$ 56111.79133 \pm 0.00011 $ & 503 ~&-0.00050& ~~~~(5) $g$ \\
$ 56111.79141 \pm 0.00013 $ & 503 ~&-0.00019& ~~~~(5) $r$ \\
$ 56111.79132 \pm 0.00013 $ & 503 ~&-0.00018& ~~~~(5) $i$ \\
$ 56111.79102 \pm 0.00013 $ & 503 ~&-0.00010& ~~~~(5) $z$ \\
$ 56130.38924 \pm 0.00042 $ & 516 ~& 0.00295& ~~~~(2) \\
$ 56194.74962 \pm 0.00027 $ & 561 ~&-0.00322& ~~~~(6) \\
$ 56217.63904 \pm 0.00013 $ & 577 ~& 0.00005& ~~~~(5) $g$ \\
$ 56217.63892 \pm 0.00011 $ & 577 ~& 0.00012& ~~~~(5) $r$ \\
$ 56217.63883 \pm 0.00010 $ & 577 ~& 0.00020& ~~~~(5) $i$ \\
$ 56217.63877 \pm 0.00012 $ & 577 ~& 0.00032& ~~~~(5) $z$ \\
$ 56227.65622 \pm 0.00062 $ & 584 ~& 0.00493& ~~~~(2) \\
$ 56407.87778 \pm 0.00014 $ & 710 ~&-0.00056& ~~~~(5) $g$ \\
$ 56407.87730 \pm 0.00033 $ & 710 ~&-0.00051& ~~~~(5) $r$ \\
$ 56407.87711 \pm 0.00017 $ & 710 ~&-0.00031& ~~~~(5) $i$ \\
$ 56407.87705 \pm 0.00021 $ & 710 ~& 0.00017& ~~~~(5) $z$ \\
$ 56460.80084 \pm 0.00016 $ & 747 ~&-0.00038& ~~~~(5) $g$ \\
$ 56460.80090 \pm 0.00020 $ & 747 ~&-0.00031& ~~~~(5) $r$ \\
$ 56460.80147 \pm 0.00028 $ & 747 ~&-0.00030& ~~~~(5) $i$ \\
$ 56460.80092 \pm 0.00025 $ & 747 ~& 0.00025& ~~~~(5) $z$ \\
$ 56510.86498 \pm 0.00013 $ & 782 ~& 0.00090& ~~~~(6) \\
$ 56510.86827 \pm 0.00067 $ & 782 ~& 0.00419& ~~~~(2) \\
$ 56520.88045 \pm 0.00067 $ & 789 ~& 0.00379& ~~~~(2) \\
$ 56533.74905 \pm 0.00013 $ & 798 ~&-0.00092& ~~~~(6) \\
$ 56882.76628 \pm 0.00065 $ &1042 ~& 0.00661& ~~~~(2) \\
$ 56942.83897 \pm 0.00083 $ &1084 ~& 0.00386& ~~~~(2) \\
\hline
\end{tabular}
\label{T0_w46}
\end{table}
The new values for the period and the reference time of mid-transit, $T_{0}$, were obtained performing a linear fit to all the mid-transit times versus their cycle number (see Tables \ref{T0_w45} and \ref{T0_w46}). We obtained:
% wasp-45
\begin{equation}
T_{0} = \mathrm{BJD(TDB)}\; 2\,455\,441.2687\,(10) + 3.1260960\,(49)\,E, \nonumber
\end{equation}
for WASP-45, and
% wasp-46
\begin{equation}
T_{0} = \mathrm{BJD(TDB)}\; 2\,455\,392.31659\,(58) +1.43036763\,(93)\,E. \nonumber
\end{equation}
for WASP-46, where the numbers in brackets represent the uncertainties on the last digit of the number they follow, and $E$ is the number of orbits the planet has completed since the $T_{0}$ used as reference. The presence of an additional planetary companion in either of the two systems can be detected thanks to the gravitational effects that it would generate on the motion of the known bodies. Indeed, if another planet orbits the same star as WASP-45\,b or WASP-46\,b, it will affect their orbital motion, by periodically advancing and retarding the transit time (e.g.\ \citealp{holman2005, lissauer2011}). 
Here, the fit has a $\chir=9.5$ and 22.7 for WASP-45 and WASP-46, respectively, indicating that the linear ephemeris is not a good match to the observations in both the cases. As noted in previous cases (e.g. \citealp{southworth2015}), this is an indication that the measured timings have too small uncertainties rather than the presence of a coherent TTV. The plots of the residuals, displayed in Figs.\,\ref{oc_w45} and \ref{oc_w46}, do not show any evidence for systematic deviations from the predicted transit times. Indeed, by fitting the residuals with both a linear and sinusoidal function, we did not find any significative correlation or periodic signal. However, more precise and homogenous measurements of the timings are mandatory to robustly establish the presence of a TTV in any of the two planetary systems.

% Figure 04
% Figure 05

\begin{figure}
\centering
\includegraphics[width=\columnwidth]{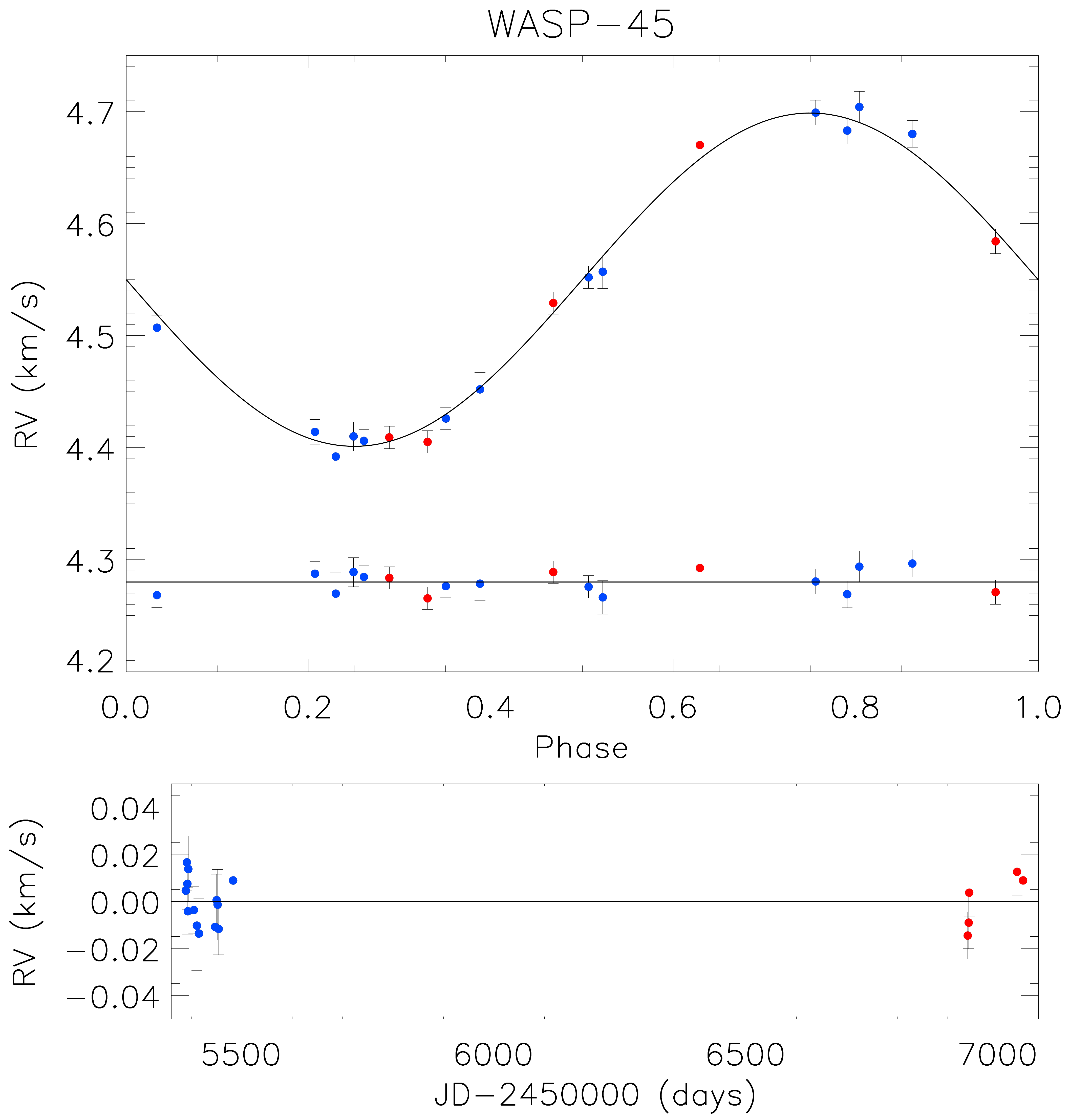}
\caption{\textit{Upper panel}: RV measurements with the best fit obtained from {\sc jktebop}; residuals are displayed at the bottom. \textit{Lower panel}: residuals of the best fit displayed as a function of the days when the spectra were observed. In both the panels blue points refer to the data from the discovery paper A12, whilst the red ones are those observed with FEROS.}
\label{RV_w45}
\end{figure}

A signature of additional planetary or more massive bodies in one of the two systems can also be found by looking for a periodicity or a linear trend in the residual of the RV data, once the sinusoidal signal due to the known planet is removed (e.g.\ \citealp{butler1999, marcy2001}). Considering both the data from the discovery paper and the new ones presented in this work, we studied the distribution of the RV residuals in time for WASP-45 (shown in Fig. \ref{RV_w45} along with the best fit), but did not find any particular trend.

% Figure 06

This is not surprising given the low probability to find a close in companion to a hot Jupiter (e.g.\ \citealp{mustill2015, ford2014, izidoro2015}).

%%%%%%%%%%%%%%%%%%%%%%%%%%%%%%%%%%%%%%%%%%%%%%%%%%%%%%
\subsection{Final photometric parameters}
\label{sec_3.2}
%%%%%%%%%%%%%%%%%%%%%%%%%%%%%%%%%%%%%%%%%%%%%%%%%%%%%%

For both the planets, each final photometric parameter was obtained as a weighted mean of the values extracted from the fit of all the individual light curves, using the relative errors as a weight. The final uncertainties, also obtained from the weighted mean, were subsequentely rescaled according to the \schir\, calculated for each quantity.
The results together with their uncertainties and the relative \chir\, are shown in Table\,\ref{ph_res}, in which they are compared with those from the discovery paper A12. The photometric parameters obtained from each single light curve are reported in  Tables \ref{pho-res_w45} and \ref{pho-res_w46} in the Appendix.

% Table 6 ------- WASP-45 & WASP-46
\begin{table*}
\caption{Photometric properties of the WASP-45 and WASP-46 systems derived by fitting the light curves with {\sc jktebop}, and taking the weighted mean of the single values obtained from each transit. The values from the discovery paper A12 are shown for comparison. $k$ is the ratio of the planetary and stellar radii, $i$ the orbital inclination, $r_{\mathrm{*}}$ and $r_{\mathrm{b}}$ the stellar and planetary relative radii respectively.}%
\centering   
\begin{tabular}{lcccccc}
\hline\hline
Parameter &        & WASP-45 &                &        & WASP-46 & \\
          & Result & \chir   & A12            & Result & \chir   & A12  \\
\hline %
$r_{\mathrm{*}}+r_{\mathrm{b}}$ &  $ 0.1172 \pm 0.0017  $ & $0.84$ &  $ 0.1217 \pm 0.0098  $    &  $ 0.1950  \pm 0.0013  $ & $1.28 $ &  $ 0.1992 \pm 0.0059 $\\ 
$k$                             &  $ 0.1095 \pm 0.0024  $ & $5.37$ &  $ 0.1234 \pm 0.0246  $    &  $ 0.14075 \pm 0.00035 $ & $1.16 $ &  $ 0.1468 \pm 0.0017 $\\ 
$i$ (deg.)                      &  $ 84.686 \pm 0.098   $ & $1.03$ &  $84.47_{-0.79}^{+0.54}$   &  $ 82.80   \pm 0.17    $ & $1.31 $ &  $ 82.63  \pm 0.38   $\\ 
$r_{\mathrm{*}}$                &  $ 0.1053 \pm 0.0014  $ & $0.98$ &  $ 0.1084 \pm 0.0094  $    &  $ 0.1709  \pm 0.0011  $ & $1.28 $ &  $ 0.1742 \pm 0.0057 $\\ 
$r_{\mathrm{b}}$                &  $ 0.01172\pm 0.00026 $ & $0.81$ &  $ 0.0134 \pm 0.0024  $    &  $ 0.02403 \pm 0.00021 $ & $1.30 $ &  $ 0.0250 \pm 0.0010 $\\ 
\hline %
\end{tabular}
\label{ph_res}%
\end{table*}

For both planets, we decided to adopt the results obtained from the fit in which we fixed the LD coefficients for all light curves. This choice was dictated by the following reasoning. In near-grazing transits, only the region near the limb of the star is transited, so there is very little information on its LD \citep{howarth2011, muller2013}. As the impact parameters of the systems are high ($b = 0.87$ for WASP-45 and $b = 0.71$ for WASP-46) and thus the transits are nearly-grazing, we decided to not fit for the LD coefficients in order to avoid biasing the results. However, we checked that the results, obtained either fixing or fitting for the LD coefficients, were compatible with each other. We assigned to the parameters of each light curve a 1-$\sigma$ uncertainty, estimated by Monte Carlo simulations, because these values were systematically higher than those obtained with the residual-permutation algorithm.

%%%%%%%%%%%%%%%%%%%%%%%%%%%%%%%%%%%%%%%%%%%%%%%%%%%%%%
\section{Physical properties of WASP-45 and WASP-46}
\label{sec_4}
%%%%%%%%%%%%%%%%%%%%%%%%%%%%%%%%%%%%%%%%%%%%%%%%%%%%%%

Using the results obtained from the photometry (Table\,\ref{ph_res}) and taking into account the spectroscopic results from the discovery paper, we redetermined the main physical parameters that characterise the two planetary systems. Following the methodology described in \cite{southworth2010}, the missing information such as the age of the system and the planetary velocity semi-amplitude were iteratively interpolated using stellar evolutionary model predictions until the best fit to the photometric and spectroscopic parameters was reached. This was done for a sequence of ages separated by 0.1\,Gyr and covering the full main sequence lifetimes of the stars. We independently repeated the interpolation using different stellar models \citep{girardi2000, claret2004,demarque2004,pietrinferni:2004,vandenberg2006, dotter2008}; for a complete list see \citealp{southworth2010}), and the final values were obtained as a weighted mean. In the final results presented in Tables \ref{fin-res_w45} and \ref{fin-res_w46}, the first uncertainty is a statistical one, which is derived by propagating the uncertainties of the input parameters, while the second is a systematic uncertainty, which takes into account the differences in the predictions coming from the different stellar models used.
The final values for the ages of the two system are not well constrained. The uncertainty that most affects the precision on these measurements is the large errorbars on $T_{\mathrm{eff}}$ (from A12 $T_{\mathrm{eff}}=5140 \pm 200$ and $T_{\mathrm{eff}}=5620 \pm 160$ for WASP-45 and WASP-46 respectively). Moreover, we noticed that for metallicity different to solar, the discrepancies between the different stellar models 
increase and therefore the systematic errorbar on the age estimation swells.

% Table 7
\begin{table}
\caption{Final results for the physical parameters of WASP-45 obtained in this work compared to those of the discovery paper. The mass $M$, radius $R$, surface gravity $g$ and mean density $\rho$ for the star and the planet are displayed; as well as the equilibrium temperature of the planet, the Safronov number $\Theta$, the semi major axis $a$ and the age of the system. $^{(a)}$This is the gyrochronological age measured in A12. The same authors also obtained a value for the stellar age from lithium abundance measurements, finding that the star is at least a few Gyr old.}
\label{fin-res_w45}
\centering
\setlength{\tabcolsep}{2pt}
\begin{tabular}{lcc}
\hline
\hline
& This work  & A12 \\
\hline
$M_{\mathrm{*}}$ ($M_{\sun}$)               & $ 0.904 \pm 0.066 \pm 0.010$   & $ 0.909  \pm 0.060       $       \\           
$R_{\mathrm{*}}$ ($R_{\sun}$)               & $ 0.917 \pm 0.024 \pm 0.003$   & $ 0.945_{-0.071}^{+0.087}$       \\           
$\log g_{\mathrm{*}}$ (cgs)                 & $ 4.470 \pm 0.014 \pm 0.002$   & $ 4.445_{-0.075}^{+0.065}$       \\           
$\rho_{\mathrm{*}}$ ($\rho_{\sun}$)         & $ 1.174 \pm 0.047          $   & $ 1.08_{-0.24}^{+0.27}   $       \\[2pt]      
$M_{\mathrm{b}}$ ($M_{\mathrm{jup}}$)       & $ 1.002 \pm 0.062 \pm 0.007$   & $ 1.007  \pm 0.053       $       \\           
$R_{\mathrm{b}}$ ($R_{\mathrm{jup}}$)       & $ 0.992 \pm 0.038 \pm 0.004$   & $ 1.16_{-0.14}^{+0.28}   $       \\           
$g_{\mathrm{b}}$ ($\mathrm{ms^{-2}}$)       & $ 25.2  \pm 1.3            $   & $ 17.0_{-6.0}^{+4.9}     $       \\           
$\rho_{\mathrm{b}}$ ($\rho_{\mathrm{jup}}$) & $ 0.959 \pm 0.077 \pm 0.003$   & $ 0.64   \pm 0.30        $       \\[2pt]      
$T_{\mathrm{eq}}$ ($\mathrm{K}$)            & $ 1170  \pm 24             $   & $ 1198 \pm 69            $       \\           
$\Theta$                                    & $ 0.0903\pm 0.0044\pm0.0003$   & $           -            $       \\           
$a$ (AU)                                    & $ 0.0405\pm 0.0010\pm0.0001$   & $0.04054 \pm 0.00090     $       \\           
Age (Gyr)                                   & $7.2_{-9.0\,-1.2}^{+5.8\,+6.8}$& $ 1.4_{-1.0}^{+2.0}      $ $^{a}$\\           
\hline
\end{tabular}
\end{table}
%
% Table 8
\begin{table}
\caption{Final results for the physical parameters of WASP-46 obtained in this work compared to those of the discovery paper. See Table. \ref{fin-res_w45} for the description of the parameters listed. This is the gyrochronological age measured in A12. $^{(a)}$The same authors also obtained a value for the stellar age from lithium abundance measurements, finding that the star is at least a few Gyr old.}
\label{fin-res_w46}
\centering
\setlength{\tabcolsep}{0pt}
\begin{tabular}{lcc}
\hline
\hline
& This work  &A12\\
\hline
$M_{\mathrm{*}}$ ($M_{\sun}$)               & $ 0.828 \pm 0.067 \pm 0.036 $    & $ 0.956 \pm 0.034 $       \\        
$R_{\mathrm{*}}$ ($R_{\sun}$)               & $ 0.858 \pm 0.024 \pm 0.013 $    & $ 0.917 \pm 0.028 $       \\        
$\log g_{\mathrm{*}}$ (cgs)                 & $ 4.489 \pm 0.013 \pm 0.006 $    & $ 4.493 \pm 0.023 $       \\        
$\rho_{\mathrm{*}}$ ($\rho_{\sun}$)         & $ 1.310 \pm 0.025           $    & $ 1.24  \pm 0.10  $       \\[2pt]   
$M_{\mathrm{b}}$ ($M_{\mathrm{jup}}$)       & $ 1.91  \pm 0.11  \pm 0.06  $    & $ 2.101 \pm 0.073 $       \\        
$R_{\mathrm{b}}$ ($R_{\mathrm{jup}}$)       & $ 1.174 \pm 0.033 \pm 0.017 $    & $ 1.310 \pm 0.051 $       \\        
$g_{\mathrm{b}}$ ($\mathrm{ms^{-2}}$)       & $ 34.3  \pm 1.1             $    & $ 28.0_{-2.0}^{+2.2}$     \\        
$\rho_{\mathrm{b}}$ ($\rho_{\mathrm{jup}}$) & $ 1.103 \pm 0.050 \pm 0.016 $    & $ 0.94  \pm 0.11  $       \\[2pt]   
$T_{\mathrm{eq}}$ ($\mathrm{K}$)            & $ 1636  \pm 44              $    & $ 1654  \pm 50    $       \\        
$\Theta$                                    & $ 0.0916\pm0.0035 \pm0.0014 $    & $  -  $                   \\        
$a$ (AU)                                    & $0.02335\pm0.00063\pm0.00034$    & $0.02448\pm0.00028$       \\        
Age (Gyr)                                   &$9.6 _{-4.2 \,-3.5}^{+3.4 \,+1.4}$&$ 1.4_{-0.6}^{+0.4} $ $^{(a)}$\\     
\hline                                                                                                               
\end{tabular}                                                                                                        
\end{table}

%%%%%%%%%%%%%%%%%%%%%%%%%%%%%%%%%%%%%%%%%%%%%%%%%%%%%%

\section{Radius vs wavelength variation}
\label{sec_5}

%%%%%%%%%%%%%%%%%%%%%%%%%%%%%%%%%%%%%%%%%%%%%%%%%%%%%%

During a transit event, a fraction of the light coming from the host star passes through the atmosphere of the planet and, according to the atmospheric composition and opacity, it can be scattered or absorbed at specific wavelengths \citep{seager2000}. Similarly to transmission spectroscopy, by observing a planetary transit at different bands simultaneously, it is then possible to look for variations in the value of the planet's radius measured in each band, and thus probe the composition of its atmosphere (e.g.\ \citealp{southworth2012,mancini2013a,narita2013}).

To pursue this goal, we phased and binned all the light curves observed with the same instrument and filter, and performed once again a fit with {\sc jktebop}. Following \cite{southworth2012}, we fixed all the parameters to the final values previously obtained (see Tables \ref{fin-res_w45} and \ref{fin-res_w46}) and fitted just for the planetary and stellar radii ratio $k$. In this way, we removed sources of uncertainty common to all datasets, maximising the relative precision of the planet/star radius ratio measurements as a function of wavelength. In order to have a set of data as homogeneous as possible, we preferred to use the light curves obtained with the same reduction pipeline and, thus, we excluded the light curves from the Euler telescope from this analysis.
%
% Figure 07
\begin{figure}
\centering
\includegraphics[width=\columnwidth]{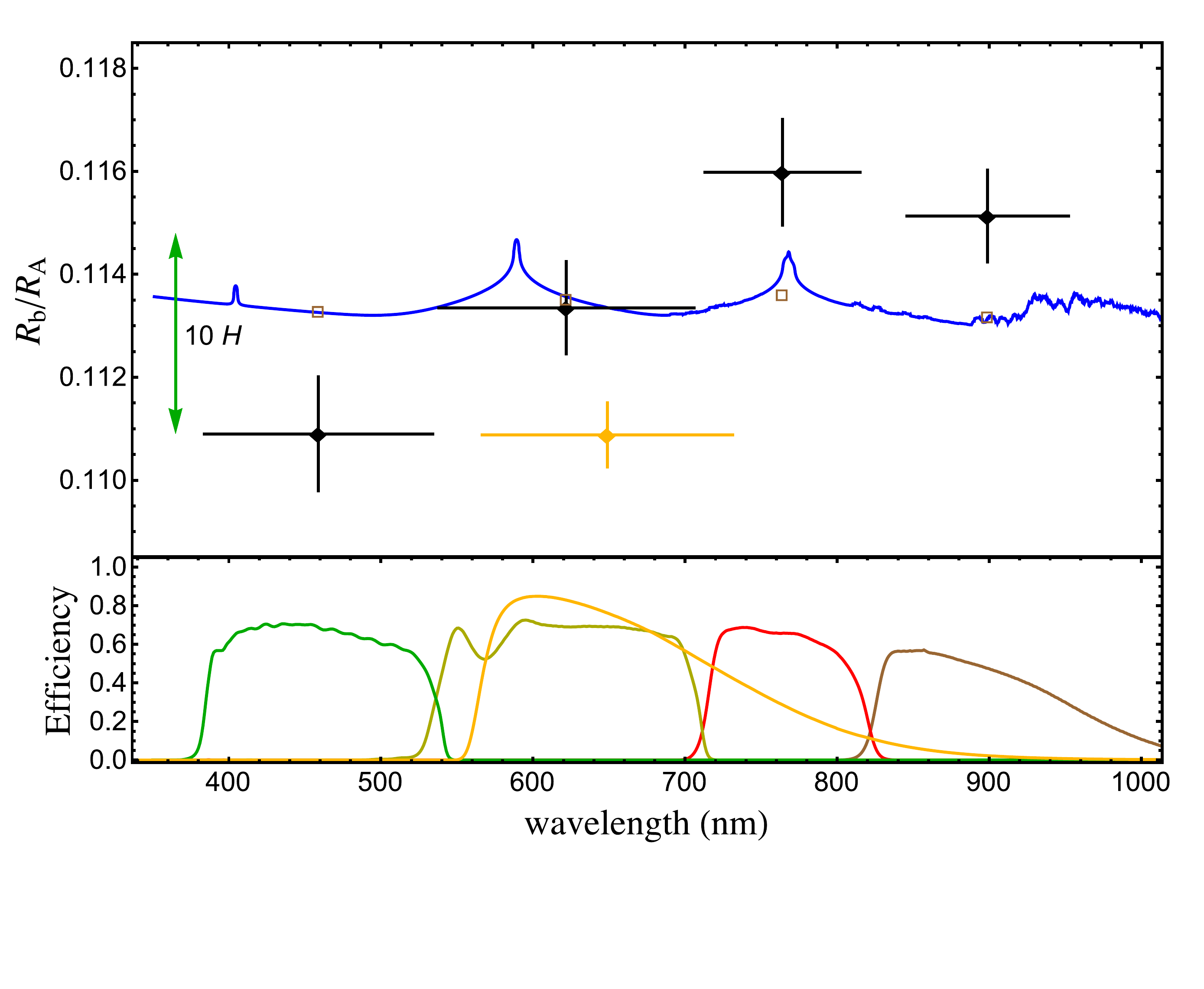}
\caption{Variation of the radius of WASP-45\,b, in terms of planet/star radius ratio, with wavelength. The black points are from the transit observed with GROND, and the yellow point is the weighted-mean results coming from the two transits observed with the Danish telescope. The vertical bars represent the uncertainties in the measurements and the horizontal bars show the FWHM transmission of the passbands used. The observational points are compared to a synthetic spectrum for a Jupiter-mass planet with a surface gravity of $g_{\mathrm{p}} = 25$\,m\,s$^{-2}$, and $T_{\mathrm{eq}}$ = 1250\,K. An offset was applied to the model to provide the best fit to our radius measurements. Transmission curves of the filters used are shown in the bottom panel. On the left of the plot the size of ten atmospheric pressure scale heights ($10\,H$) is shown. The small coloured squares represent band-averaged model radii over the bandpasses used in the observations.}
\label{w45_wl-r}
\end{figure}%
%
% Figure 08
\begin{figure}
\centering
\includegraphics[width=9.1cm]{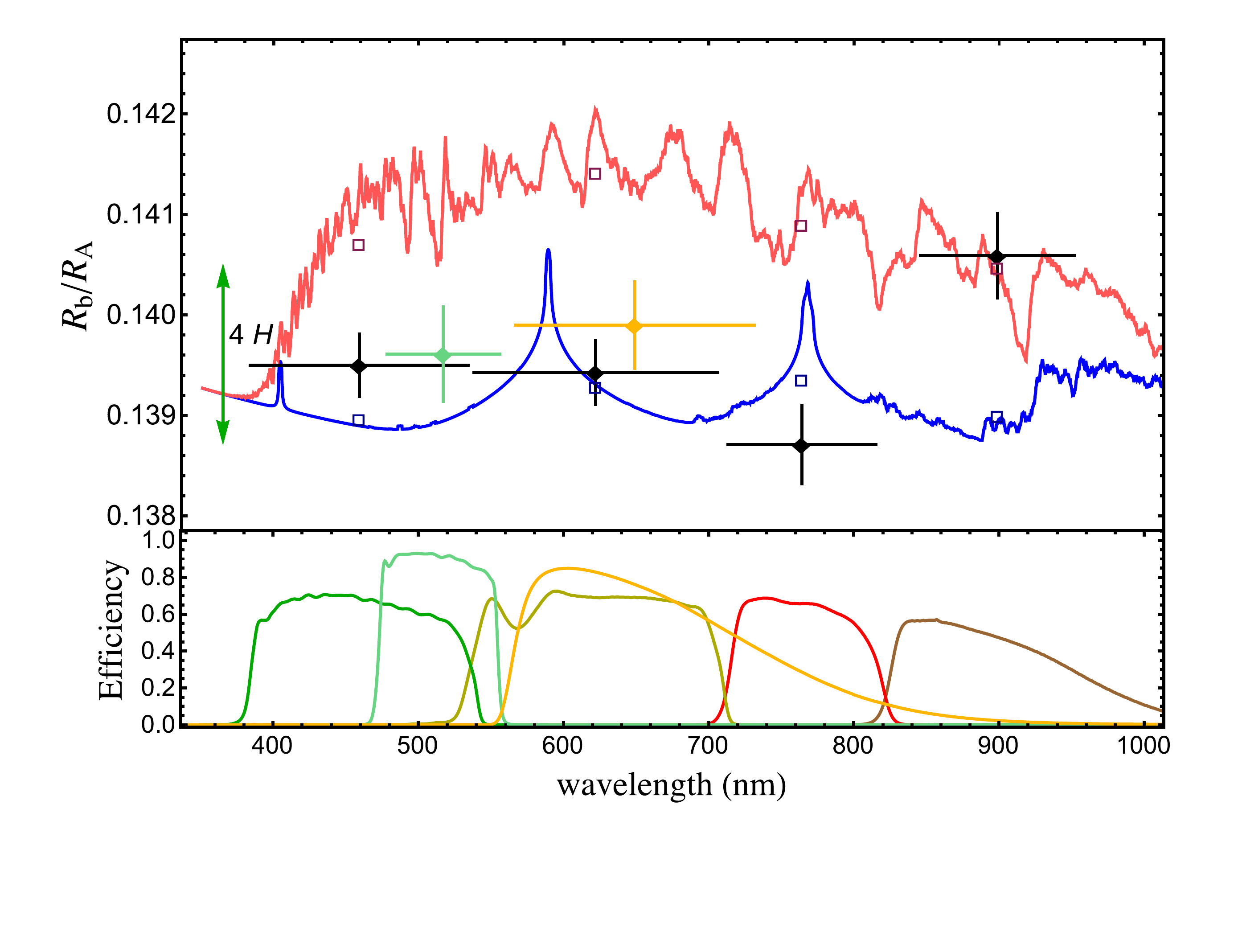}
\caption{Variation of the radius of WASP-46\,b, in terms of planet/star radius ratio, with wavelength. The black points are from the transit observed with GROND, the green point is that obtained with the NTT, and the yellow point is the weighted-mean result coming from the three transits observed with the Danish telescope. The vertical bars represent the uncertainties in the measurements and the horizontal bars show the FWHM transmission of the passbands used. The observational points are compared to two synthetic spectra for a Jupiter-mass planet with a surface gravity of $g_{\mathrm{p}} = 25$\,m\,s$^{-2}$, and $T_{\mathrm{eq}} = 1750$\,K. The synthetic spectrum in blue does not include TiO and VO opacity, while the spectrum in red does, based on equilibrium chemistry. An offset was applied to the models to provide the best fit to our radius measurements. Transmission curves of the filters used are shown in the bottom panel. On the left of the plot the size of four atmospheric pressure scale heights ($4\,H$) is shown. The small coloured squares represent band-averaged model radii over the bandpasses used in the observations.}
\label{w46_wl-r}
\end{figure}%
%

% Table 9
\begin{table}
\caption{
Values of $k$ for each of the light curves as plotted in
Figs.\,\ref{w45_wl-r} and \ref{w46_wl-r}.}
\centering
\setlength{\tabcolsep}{2pt}
\begin{tabular}{lccc}
\hline
\hline
Passband & Central  & FWHM & $k$ \\
             & wavelength (nm)  & (nm) &  \\
\hline\\[-6pt]
\multicolumn{2}{l}{\textbf{WASP-45:}}              \\[2pt]%
GROND $g^{\prime}$  & 477.0 & 137.9 & $ 0.11090 \pm 0.00111 $      \\
GROND $r^{\prime}$  & 623.1 & 138.2 & $ 0.11335 \pm 0.00090 $       \\
Bessel $R$ & 648.9 & 164.7 & $0.11088 \pm 0.00062$     \\
Gunn $r$ & 664.1 & \,\,85.0 & $ 0.10821 \pm 0.00099$ \\
GROND $i^{\prime}$  & 762.5 & 153.5 & $ 0.11598 \pm 0.00103 $      \\
GROND $z^{\prime}$  & 913.4 & 137.0 & $ 0.11513 \pm 0.00089 $       \\[3pt]%
\multicolumn{2}{l}{\textbf{WASP-46:}}              \\[2pt]%
GROND $g^{\prime}$  & 477.0 & 137.9 & $ 0.13950 \pm 0.00031 $      \\
Gunn $g$ & 516.9 & \,\,77.6 & $0.13961 \pm 0.00047$ \\
GROND $r^{\prime}$  & 623.1 & 138.2 & $ 0.13943 \pm 0.00032 $       \\
Bessel $R$ & 648.9 & 164.7 & $0.13990 \pm 0.00043$     \\
Gunn $r$ & 664.1 & \,\,85.0 & $ 0.13815 \pm 0.00113$ \\
GROND $i^{\prime}$  & 762.5 & 153.5 & $ 0.13871 \pm 0.00039 $      \\
GROND $z^{\prime}$  & 913.4 & 137.0 & $ 0.14059 \pm 0.00042 $       \\
\hline
\end{tabular}
\label{tab_k}
\end{table}

The values of $k$ that we obtained at different passbands are reported in Table\,\ref{tab_k} and illustrated in Figs.\ \ref{w45_wl-r} and \ref{w46_wl-r} for WASP-45 and WASP-46, respectively. In these figures, for comparison, we also show the expected values of the planetary radius in function of wavelength, obtained from synthetic spectra constructed from model planetary atmospheres by \cite{fortney2010}, using different molecular compositions. The models were estimated for a Jupiter-mass planet with a surface gravity of $g_{\mathrm{p}} = 25$\,m\,s$^{-2}$, a base radius of 1.25\,$R_{\mathrm{Jup}}$ at 10\,bar, and $T_{\mathrm{eq}} = 1250$\,K and 1750\,K for WASP-45\,b and WASP-46\,b, respectively. The model displayed with a red line in Fig.\,\ref{w46_wl-r} was run in an isothermal case taking into account chemical equilibrium and the presence of strong absorbers, such as TiO and VO. The models displayed with blue lines in Figs.\ \ref{w45_wl-r} and \ref{w46_wl-r} were obtained omitting the presence of the metal oxides.

Looking at the distribution of the experimental points in the two figures, we do not see the telltale increase of the radius at the shortest wavelengths (e.g.\ see \citealp{lecavelierdesetangs2008}), and therefore we do not expect a strong Rayleigh scattering in the atmosphere.
However by studying our data points quantitatively we can not exclude any hypothesis. By performing a Monte Carlo simulation, we obtained that our data points are consistent within 3$\sigma$ to a slope with a maximum inclination of $m= -1.40\times10^{-5}$ for WASP-45\,b and $m= -1.17\times10^{-5}$ for WASP-46\,b (where with $m$ we indicate the slope coefficient of the best linear fit). Fitting with a straight line the predictions given at short wavelengths by a model with the Rayleigh scattering enhanced by a factor of 1000, we obtained slope coefficients lower that the ones just mentioned (the slope coefficient is $m= -2.2\times10^{-6}$, and $m= -3.3\times10^{-6}$ for WASP-45 and WASP-46, respectively). Although pointing in the direction of no strong Rayleigh scattering, our data are not sufficient to completely rule out this scenario. More data points are needed to make stronger statements regarding this matter.

For the case of WASP-45\,b, for which we have only one transit observed with GROND, it is possible to note a radius variation between the $g^{\prime}$ and $i^{\prime}$ bands at $2 \sigma$, corresponding to roughly 12$H$ pressure scale heights (where the atmospheric pressure scale height is $H = k\,\mathrm{T}_{eq}/\mu \, g$, with $k$ being the Boltzmann constant, T$_{eq}$ the planetary atmosphere temperature, $\mu$ the mean molecular weight and $g$ the planet's surface gravity). In the case of WASP-46\,b, for which we observed four transits with GROND, we noticed a small variation of $\sim 4\,H$ between the $i^{\prime}$ and $z^{\prime}$ bands but at only $1.5 \sigma$. These detections are too small to be significant -- both planets are not well suited to transmission photometry or spectroscopy due to their large impact parameters and high surface gravities.
%
% Figure 09
\begin{figure*}
\centering
\includegraphics[width=13.0cm]{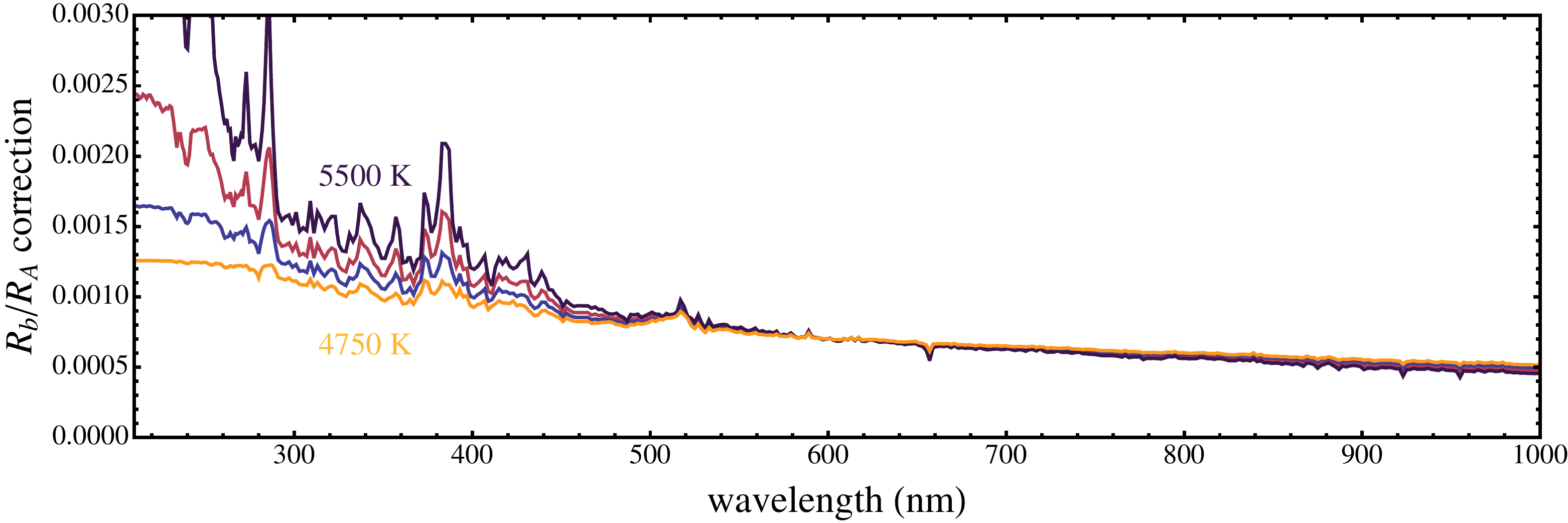}
\caption{The effect of the presence of unocculted starspot on the surface of the star on the transmission spectrum, considering a 1 per cent flux drop at 600nm. The stellar temperature adopted is $T_{\mathrm{eff}}= 5600$\,K, and the spots coverage is modelled using a grid of stellar atmospheric models of different temperature ranging from 4750 (yellow line) to 5500 (purple line), with steps of 250\,K.}
\label{corrections}
\end{figure*}%

As stated in A12, the lightcurve of WASP-46 shows a rotational modulation, which is symptomatic of stellar activity. The presence of star spots on the stellar surface, and in general stellar activity, can produce variations in the transit depth when it is measured at different epochs. In particular, we expected that such a variation is stronger at bluer wavelengths and affecting more the lightcurves obtained through the $g^{\prime}$ band, whereas it is negligible in the $i$ and $z$ bands (e.g., \citealp{sing2011,mancini2014}). Correcting for this effect, would slightly shift the data point relative to the bluer bands, towards the bottom of Fig. \ref{w45_wl-r}. Anyway, since the stellar activity is not particularly high, the expected variation in the transit depth is small and within our errorbars (Fig. \ref{corrections} shows the effect of the presence of starspots, at different temperatures, on the transit depth with wavelength. The stellar model used to produce the curves are the ATLAS9 by \citealp{castelli2004}). 

%
%%%%%%%%%%%%%%%%%%%%%%%%%%%%%%%%%%%%%%%%%%%%%%%%%%%%%%
\section{Summary and Conclusions}
\label{sec_6}
%%%%%%%%%%%%%%%%%%%%%%%%%%%%%%%%%%%%%%%%%%%%%%%%%%%%%%

% Figure 10
\begin{figure*}
\centering
\includegraphics[width=\textwidth]{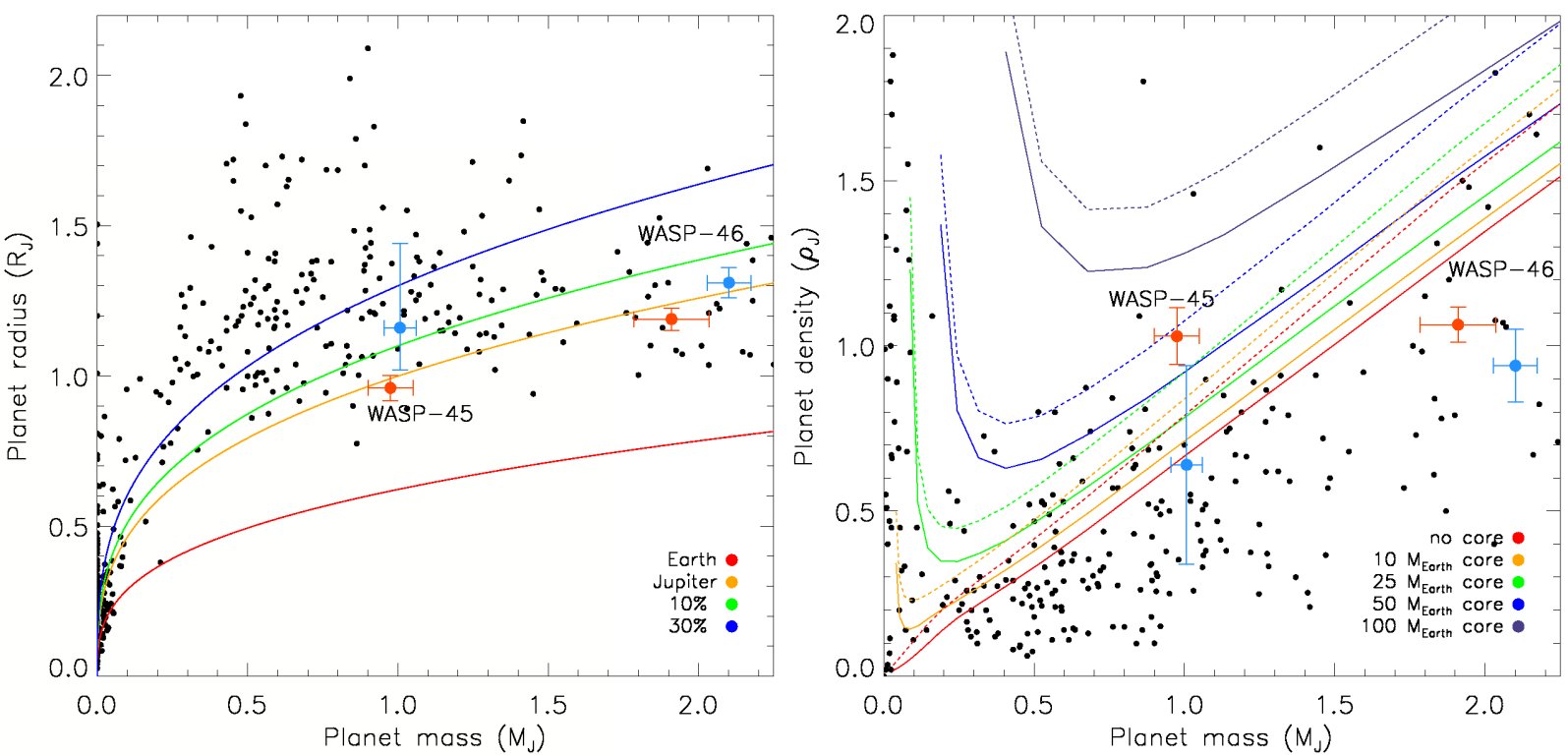}
\caption{\textit{Left}: the mass versus radius diagram of the known transiting planets. The values for WASP-45\,b and WASP-46\,b obtained in this work are displayed in orange, and for comparison we also show in light blue the measurements reported in the discovery paper. The coloured curves represent the iso-density lines for planets with the density of Earth and Jupiter, and with a density equal to a Jupiter-like planet with a radius inflated by 10\% and 30\%.
\textit{Right}: the mass versus density diagram of the known transiting planets. As in the left panel, the values obtained for WASP-45 and WASP-46 are highlighted in colour. The superimposed lines represent the expected radius of the planet having an inner core of 0, 10, 25, 50 and 100 Earth masses, and calculated for 10\,Gyr old planets at 0.02\,au, solid lines, and 0.045\,au, dashed lines \citep{fortney2007}.}
\label{r-m-rho}
\end{figure*}

We have presented new multi-band photometric light curves of transit events of the hot-Jupiter planets WASP-45\,b and WASP-46\,b, and new RV measurements of WASP-45. We used these new datasets to refine the orbital and physical parameters that characterise the architecture of the WASP-45 and WASP-46 planetary systems. Moreover, we used the light curves observed through several optical passbands to probe the atmosphere of the two planets. Our conclusions are as follows:

\begin{itemize}

\item [$\bullet$]
The radius of WASP-45\,b is now much better constrained, with a precision better by almost one order of magnitude with respect to that of A12. We found that WASP-46\,b has a slightly lower radius ($1.189 \pm 0.037\, R_{\mathrm{Jup}}$) compared to the previous estimation ($1.310 \pm 0.051\, R_{\mathrm{Jup}}$). The left panel of Fig.\,\ref{r-m-rho} shows the change in position in the planet mass-radius diagram for both WASP-45\,b and WASP-46\,b. \\

\item [$\bullet$]
Based on our estimates, both planets have a larger density than previously thought (see the right panel of Fig.\,\ref{r-m-rho}). In particular, WASP-45\,b appears to be 
one of the densest planets in its mass regime (there are only 3 other planets with masses between $0.7$ and $1.3 M_{\mathrm{Jup}}$ that have similar or higher density), suggesting the presence of a heavy-element core of roughly $50\,M_{\oplus}$ \citep{fortney2007}. \\

\item [$\bullet$]
By studying the transit times and the RV residuals we did not find any hint for the presence of any additional planetary companion in either of the two planetary systems. However more spectroscopic and photometric data are necessary to claim  a lack of other planetary companions, at any mass and separation, in these two systems. In particular a higher temporal cadence for the photometric observations could allow to detect the signature of smaller inner planets, while RV measurements obtained at different epochs separated by several months/years could provide information on the presence of outer long period planets. \\

\item [$\bullet$]
Looking at the radius variation in terms of wavelength, we estimated the upper limit of the slope allowed by our data within $3 \sigma$. The slope of the best linear fit are $m= -1.40\times10^{-5}$ and $m= -1.17\times10^{-5}$ for WASP-45\,b and WASP-46\,b respectively. Comparing these values to the slope obtained from a model with 1000 times enhancement of Rayleigh scattering we found that we can not exclude with a high statistical significance the presence of strong Rayleigh scattering in the atmospheres of both planets.

The data of one transit of WASP-45\,b, observed simultaneously in four optical bands with GROND, indicate a planetary radius variation of more than $10\,H$ between the $g^{\prime}$ and the $i^{\prime}$ bands, but at only a $\sim 2\sigma$ significance level. Such a variation is rather high for that expected for a planet with a temperature below 1200\,K, and would require the presence of strong absorbers between 800 and 900\,nm. More observations are requested to verify this possible scenario.

In the case of WASP-46\,b, by joining the GROND multi-colour data of four transit events, we detected a very small radius variation, roughly $4\,H$, between the $i^{\prime}$ and $z^{\prime}$ bands, but at only a $1.5\sigma$ significance level. This variation can be explained by supposing the absence of potassium at 770\,nm and a significant amount of water vapour around 920\,nm.

\end{itemize}

\section*{Acknowledgments}

We acknowledge the use of the NASA Astrophysics Data System; the SIMBAD database operated at CDS, Strasbourg, France; and the arXiv scientific paper preprint service operated by Cornell University. This work was supported by KASI (Korea Astronomy and Space Science Institute) grants 2012-1-410-02 and 2013-9-400-00. ASB acknowledges support from the European Union Seventh Framework Programme (FP7/2007-2013) under grant agreement number 313014 (ETAEARTH). TCH acknowledges financial support from the Korea Research Council for Fundamental Science and Technology (KRCF) through the Young Research Scientist Fellowship Program. YD, AE, JSurdej and OW acknowledge support from the Communaut\'{e} fran\c{c}aise de Belgique -- Actions de recherche concert\'{e}es -- Acad\'{e}mie Wallonie-Europe. SHG and XBW would like to thank the financial support from National Natural Science Foundation of China through grants Nos. 10873031 and 11473066.
SC thanks G-D. Marleau for useful discussion and comments, the staff and astronomers observing at the ESO La Silla observatory during January and February 2015 for the great, friendly and scientifically stimulating environment.

\appendix

\section{Photometric parameters}
\label{appendix}
In the two tables in this appendix are presented the photometric results obtained with {\sc jktebop} from the fit of each light curve presented in the paper.

% Table A1 *** WASP-45
\begin{table*}
\centering
\begin{tabular}{lccccc}
\hline\hline
Source & $r_{A}+r_{b}$& $k$ & $i$ (deg.) & $r_{A}$ &$r_{b}$\\
\hline %
Eul\,1.2m                & $0.1218 \pm 0.0044$ & $0.1018 \pm 0.0018$ & $84.3258 \pm 0.2408$ & $0.11052 \pm 0.0039$ & $0.011255 \pm 0.00053$ \\ %    
Dan\,1.54m (transit \#1) & $0.1188 \pm 0.0032$ & $0.1130 \pm 0.0022$ & $84.6423 \pm 0.1809$ & $0.10671 \pm 0.0027$ & $0.012061 \pm 0.00049$ \\ %    
MPG\,2.2m  (transit g)   & $0.1165 \pm 0.0057$ & $0.1155 \pm 0.0063$ & $84.7024 \pm 0.3381$ & $0.10444 \pm 0.0046$ & $0.012060 \pm 0.00116$ \\ %    
MPG\,2.2m  (transit r)   & $0.1140 \pm 0.0036$ & $0.1128 \pm 0.0022$ & $84.9141 \pm 0.2079$ & $0.10245 \pm 0.0030$ & $0.011558 \pm 0.00054$ \\ %    
MPG\,2.2m  (transit i)   & $0.1181 \pm 0.0061$ & $0.1234 \pm 0.0089$ & $84.6154 \pm 0.3761$ & $0.10516 \pm 0.0046$ & $0.012979 \pm 0.00149$ \\ %    
MPG\,2.2m  (transit z)   & $0.1198 \pm 0.0046$ & $0.1176 \pm 0.0035$ & $84.5609 \pm 0.2778$ & $0.10719 \pm 0.0038$ & $0.012601 \pm 0.00079$ \\ %
Dan\,1.54m (transit \#2) & $0.1076 \pm 0.0060$ & $0.1097 \pm 0.0035$ & $85.2537 \pm 0.3657$ & $0.09699 \pm 0.0051$ & $0.010640 \pm 0.00088$ \\ %
\hline %

{\bf Final results} &   $\mathbf{0.1172\pm0.0017}$&$\mathbf{0.1095\pm0.0024}$&$\mathbf{84.686\pm0.098}$& $\mathbf{0.1053\pm0.0014}$& $\mathbf{0.01172\pm0.00026}$\\%
\hline
\citet{anderson2012}& $0.1217 \pm 0.0098$ & $0.1234 \pm 0.0246$ & $84.47_{-0.79}^{+0.54}$ & $0.1084 \pm 0.0094$ & $0.0134 \pm 0.0024$  \\ %
\hline
\end{tabular}
\caption{Photometric properties of the WASP-45 system derived by fitting the light curves with {\sc jktebop}. In bold are highlighted the final parameters obtained as weighted mean. The values from the discovery paper are also shown for comparison.}%
\label{pho-res_w45}%
\end{table*}

% Table A2 *** WASP-46
\begin{table*}
\centering
\begin{tabular}{lccccc}
\hline\hline
Source & $r_{A}+r_{b}$& $k$ & $i$ (deg.) & $r_{A}$ &$r_{b}$\\
\hline %
Eul\,1.2m  (transit \#1)   & $0.1964 \pm 0.0190$ & $0.13551 \pm 0.00585$ & $82.90 \pm 1.31$ & $0.1730 \pm 0.0160$ & $0.02344 \pm 0.00302$ \\ %
Eul\,1.2m  (transit \#2)   & $0.1857 \pm 0.0099$ & $0.14025 \pm 0.00296$ & $83.51 \pm 0.70$ & $0.1629 \pm 0.0084$ & $0.02284 \pm 0.00158$ \\ %
NTT\,3.58m                 & $0.1979 \pm 0.0044$ & $0.14123 \pm 0.00139$ & $82.84 \pm 0.30$ & $0.1734 \pm 0.0036$ & $0.02449 \pm 0.00074$ \\ %
MPG\,2.2m  (transit \#1 g) & $0.1996 \pm 0.0041$ & $0.14226 \pm 0.00127$ & $82.72 \pm 0.27$ & $0.1747 \pm 0.0034$ & $0.02485 \pm 0.00068$ \\ %
MPG\,2.2m  (transit \#1 r) & $0.1997 \pm 0.0046$ & $0.14229 \pm 0.00115$ & $82.77 \pm 0.31$ & $0.1748 \pm 0.0039$ & $0.02487 \pm 0.00072$ \\ %
MPG\,2.2m  (transit \#1 i) & $0.1938 \pm 0.0042$ & $0.13991 \pm 0.00102$ & $83.05 \pm 0.30$ & $0.1700 \pm 0.0035$ & $0.02379 \pm 0.00063$ \\ %
MPG\,2.2m  (transit \#1 z) & $0.1925 \pm 0.0069$ & $0.14051 \pm 0.00119$ & $83.34 \pm 0.49$ & $0.1688 \pm 0.0059$ & $0.02372 \pm 0.00100$ \\ %
Dan\,1.54m (transit \#1)   & $0.1901 \pm 0.0095$ & $0.14291 \pm 0.00310$ & $83.40 \pm 0.69$ & $0.1664 \pm 0.0080$ & $0.02378 \pm 0.00157$ \\ %
MPG\,2.2m  (transit \#2 g) & $0.1971 \pm 0.0048$ & $0.14185 \pm 0.00162$ & $82.85 \pm 0.33$ & $0.1726 \pm 0.0040$ & $0.02449 \pm 0.00082$ \\ %
MPG\,2.2m  (transit \#2 r) & $0.1902 \pm 0.0037$ & $0.13955 \pm 0.00108$ & $83.25 \pm 0.25$ & $0.1669 \pm 0.0031$ & $0.02329 \pm 0.00059$ \\ %
MPG\,2.2m  (transit \#2 i) & $0.1968 \pm 0.0042$ & $0.14120 \pm 0.00117$ & $82.93 \pm 0.29$ & $0.1725 \pm 0.0036$ & $0.02435 \pm 0.00066$ \\ %
MPG\,2.2m  (transit \#2 z) & $0.1953 \pm 0.0041$ & $0.14135 \pm 0.00108$ & $82.96 \pm 0.28$ & $0.1711 \pm 0.0035$ & $0.02418 \pm 0.00064$ \\ %
MPG\,2.2m  (transit \#3 g) & $0.1805 \pm 0.0103$ & $0.13800 \pm 0.00288$ & $84.14 \pm 0.80$ & $0.1586 \pm 0.0087$ & $0.02189 \pm 0.00163$ \\ %
MPG\,2.2m  (transit \#3 r) & $0.2067 \pm 0.0053$ & $0.14525 \pm 0.00169$ & $82.25 \pm 0.33$ & $0.1805 \pm 0.0044$ & $0.02621 \pm 0.00091$ \\ %     
MPG\,2.2m  (transit \#3 i) & $0.1562 \pm 0.0149$ & $0.13543 \pm 0.00370$ & $86.30 \pm 1.69$ & $0.1376 \pm 0.0127$ & $0.01863 \pm 0.00213$ \\ %     
MPG\,2.2m  (transit \#3 z) & $0.1723 \pm 0.0093$ & $0.13828 \pm 0.00177$ & $84.72 \pm 0.76$ & $0.1514 \pm 0.0080$ & $0.02093 \pm 0.00130$ \\ %     
MPG\,2.2m  (transit \#4 g) & $0.1971 \pm 0.0061$ & $0.13918 \pm 0.00209$ & $82.91 \pm 0.41$ & $0.1730 \pm 0.0051$ & $0.02408 \pm 0.00103$ \\ %     
MPG\,2.2m  (transit \#4 r) & $0.1958 \pm 0.0149$ & $0.14121 \pm 0.00334$ & $82.79 \pm 1.01$ & $0.1716 \pm 0.0126$ & $0.02423 \pm 0.00231$ \\ %     
MPG\,2.2m  (transit \#4 i) & $0.1789 \pm 0.0113$ & $0.13713 \pm 0.00217$ & $84.03 \pm 0.82$ & $0.1574 \pm 0.0097$ & $0.02158 \pm 0.00161$ \\ %
MPG\,2.2m  (transit \#4 z) & $0.1945 \pm 0.0090$ & $0.14015 \pm 0.00298$ & $83.30 \pm 0.69$ & $0.1706 \pm 0.0075$ & $0.02391 \pm 0.00149$ \\ %
Dan\,1.54m (transit \#2)   & $0.1959 \pm 0.0046$ & $0.14088 \pm 0.00153$ & $82.98 \pm 0.31$ & $0.1717 \pm 0.0038$ & $0.02419 \pm 0.00076$ \\ %
Dan\,1.54m (transit \#3)   & $0.1930 \pm 0.0049$ & $0.14039 \pm 0.00136$ & $83.09 \pm 0.34$ & $0.1692 \pm 0.0042$ & $0.02376 \pm 0.00078$ \\ %
\hline %
{\bf Final results}        &$\mathbf{0.1950\pm0.0013}$&$\mathbf{0.14075\pm0.00035}$&$\mathbf{82.80\pm0.17}$&$\mathbf{0.1709\pm0.0011}$&$\mathbf{0.02403\pm0.00021}$\\%
\hline
\citet{anderson2012}       & $0.1992 \pm 0.0059$ & $0.1468  \pm 0.0017$  & $82.63  \pm 0.38 $ & $0.1742 \pm 0.0057$ & $0.0250  \pm 0.0010 $ \\ %
\hline
\end{tabular}
\caption{Photometric properties of the WASP-46 system derived by fitting the light curves with {\sc jktebop}. In bold are highlighted the final parameters obtained as weighted mean. The values from the discovery paper are also shown for comparison.}%
\label{pho-res_w46}%
\end{table*}

\bsp

\label{lastpage}

\end{document}